\documentclass[aps,preprint,amsmath,amssymb,amsfonts]{revtex4}
\usepackage{epsfig}
\usepackage{graphicx}
\usepackage{subfigure}
\usepackage{dcolumn}
\usepackage{bm}
\usepackage{amsthm}
\usepackage{enumitem}
\usepackage{slashed}
\usepackage{braket}
\usepackage{amsmath}
\usepackage{mathtools}
\usepackage{bbold}

\usepackage{color}   
\usepackage{hyperref}
\hypersetup{
	colorlinks=true,  
	linkcolor=black,  
	citecolor=black,
	filecolor=black,
	urlcolor=black,
}

\makeatletter
\def\l@subsubsection#1#2{}
\makeatother

\newcommand{\const}{\mbox{const}}
\newcommand{\del}{\partial}

\renewcommand{\Im}{\operatorname{Im}}

\newcommand{\tr}{\operatorname{tr}}

\newcommand{\diag}{\operatorname{diag}}

\newcommand{\bbC}{\mathbb{C}}

\newcommand{\bbN}{\mathbb{N}}
\newcommand{\bbP}{\mathbb{P}}
\newcommand{\bbR}{\mathbb{R}}
\newcommand{\bbT}{\mathbb{T}}
\newcommand{\bbZ}{\mathbb{Z}}

\newcommand{\calB}{\mathcal{B}}

\newcommand{\calJ}{\mathcal{J}}
\newcommand{\calK}{\mathcal{K}}

\newcommand{\calN}{\mathcal{N}}
\newcommand{\calO}{\mathcal{O}}

\newcommand{\scri}{\mathcal{I}}

\begin{document}

\title{Higher-Spin dS/CFT beyond minimal type-A,\\and the doubled nature of de Sitter space}

\author{Julian Lang}
\email{julian.lang.research@gmail.com}
\affiliation{Okinawa Institute of Science and Technology, 1919-1 Tancha, Onna-son, Okinawa 904-0495, Japan}

\author{Yasha Neiman}
\email{yashula@icloud.com}
\affiliation{Okinawa Institute of Science and Technology, 1919-1 Tancha, Onna-son, Okinawa 904-0495, Japan}

\date{\today}

\begin{abstract}
We study the holographic duality between free vector models and Higher-Spin Gravity in 4-dimensional de Sitter space. On both the bulk and boundary sides, we fix reality conditions which allow us to include odd spins in the bulk, and to extend the duality from type-A to type-B (i.e. from boundary scalars to spinors). The bulk reality conditions have been considered before, and are closely related to a higher-spin-symmetric Hilbert-space norm for the single-particle (super)multiplet in de Sitter. The boundary reality conditions are new. After thus extending higher-spin dS/CFT, we construct two alternative viewpoints on the type-A and type-B theories. In one view, the two co-exist in ``parallel universes'', specifically in even vs. odd antipodal identifications of de Sitter space. In the other view, type-A and type-B are Legendre-transformed descriptions of the same twistor-space theory in Euclidean AdS (this involves the observation that the higher-spin partition function is Legendre-dual to itself). In this way, higher-spin holography reconciles the two-boundary nature of de Sitter space with the one-boundary nature of Euclidean AdS. The supersymmetric theory that unifies type-A and type-B with bulk fermions fails to be consistent under the appropriate reality conditions. 
\end{abstract}

\maketitle
\tableofcontents
\newpage

\section{Introduction} \label{sec:intro}

\subsection{Higher-Spin Gravity, dS/CFT and twistors}

Higher-Spin (HS) Gravity \cite{Vasiliev:1990en,Vasiliev:1992av,Vasiliev:1995dn,Vasiliev:1999ba} is the putative interacting theory of an infinite multiplet of massless fields with increasing spins, in 4d spacetime with cosmological constant. It is a cousin of supergravity and string theory, with its own AdS/CFT correspondence \cite{Klebanov:2002ja,Sezgin:2002rt,Sezgin:2003pt,Giombi:2012ms}. While certain extended versions of HS Gravity are dual to string theory \cite{Chang:2012kt}, Vasiliev's original proposal \cite{Vasiliev:1990en,Vasiliev:1992av} seems to be a genuinely distinct, smaller theory. In this paper, we consider this theory and several of its sub-sectors. In AdS, these theories and their holographic duals are as follows:
\begin{enumerate}
	\item The original theory \cite{Vasiliev:1990en,Vasiliev:1992av} is $\calN=2$ supersymmetric, with two fields of every (integer and half-integer) spin. Its AdS/CFT boundary dual (with the simplest boundary conditions) is a free $U(N)$ vector model of complex scalar fields $\phi^I,\bar\phi_I$ and their spinor superpartners $\psi^I,\bar\psi_I$ (here, $I$ is a color index, and spinor indices are suppressed). 
	\item There is a consistent truncation to the bosonic sector. This consists of two subsectors that don't interact with each other, known as the type-A and type-B theories. Each has one field of every integer spin. On the boundary, they correspond to the $U(N)$ vector model of only scalars $\phi^I,\bar\phi_I$ or only spinors $\psi^I,\bar\psi_I$, respectively.
	\item There is a separate consistent truncation to ``minimal'' theories. On the boundary, this involves treating each complex field $\phi^I$ or $\psi^I$ as a pair of real fields, enlarging the color symmetry from $U(N)$ to $O(2N)$ (we disregard odd orthogonal groups $O(2N+1)$, which won't be relevant to the de Sitter case). When applied to the $\calN=2$ theory, this truncation leads to a smaller, $\calN=1$ supersymmetric model \cite{Engquist:2002vr}. When applied to the type-A and type-B bosonic sectors, it restricts them to even spins.
\end{enumerate}
In many ways, HS Gravity (without extensions or strong coupling as in \cite{Chang:2012kt}) is less realistic than string theory. For instance, already the cubic self-interactions of the spin-2 field are not those of GR, but contain a curvature-cubed term scaled by the cosmological radius (unlike in string theory, where higher-derivative couplings are suppressed by the string length). Nevertheless, HS Gravity agrees with the real world in two key features: it is native to 4d spacetime, and it admits a \emph{positive} cosmological constant. While the original Vasiliev equations \cite{Vasiliev:1990en,Vasiliev:1992av} were written in Lorentzian AdS, they've been adapted to other signatures in \cite{Iazeolla:2007wt,Chang:2013afa} (bosonic) and \cite{Sezgin:2012ag} (supersymmetric). More importantly, it was observed in \cite{Anninos:2011ui} that, for the minimal type-A model, one can turn the AdS/CFT correspondence into dS/CFT in the sense of \cite{Maldacena:2002vr}, \emph{with the correct sign of the 2-point functions}, by flipping the statistics of the boundary scalars $\phi^I$ from Bose to Fermi, and changing the color group from $O(2N)$ to $Sp(2N)$. Later, a difficulty was noted \cite{Anninos:2012ft}: already for a constant scalar source on the $S_3$ boundary of de Sitter, the CFT partition function has a \emph{local} maximum at zero (as ensured by the 2-point function), but not a \emph{global} one. A solution was proposed in \cite{David:2020ptn}: the CFT partition function can be replaced by a HS-algebraic expression, which agrees with it on all correlators at separated points, but differs at finite sources, due to a subtlety involving sign ambiguities that can't be consistently resolved (see \cite{David:2020ptn} for much more detail). While we still find this solution conceptually confusing, it has a pair of clear virtues: the partition function is manifestly HS-symmetric, and it exhibits a global maximum at constant scalar sources. In this paper, when considering the partition function at finite sources, we will follow the proposal of \cite{David:2020ptn}. 

HS Gravity has been treated in a number of formalisms, with different attitudes towards locality, background spacetime, and global/local HS symmetries. The main formalism we'll use in this paper is that of \cite{Neiman:2017mel,David:2020ptn}, which is a manifestly (A)dS-covariant rewriting of \cite{Didenko:2012tv,Didenko:2013bj}. In this language, boundary operators are mapped to twistor functions $F(Y)$. These in turn correspond (via the Penrose transform) to linearized bulk fields, packaged as master fields $C(x;Y)$ that contain gauge-invariant curvatures of all spins along with infinite towers of derivatives. The formalism has manifest global HS symmetry, expressed via the HS algebra on twistors $Y$. Boundary correlators are expressed as HS-algebraic traced products, either of the bulk master fields $C(x;Y)$ \cite{Didenko:2012tv,Didenko:2013bj}, or of the twistor functions $F(Y)$ \cite{Neiman:2017mel,David:2020ptn}. It is this HS-algebraic construction that, when ``taken seriously'', leads to the better-behaved partition function mentioned above.

\subsection{Questions addressed in this paper}

Our main hope concerning Higher-Spin dS/CFT is that, as a working model, it can shed light on general questions about quantum gravity in de Sitter space. The results of this paper revolve around three such questions, two more technical and one more conceptual:
\begin{enumerate}
	\item Can dS/CFT be supersymmetric? In HS Gravity, we re-affirm that the answer is ``no'', for the same reason as in ordinary de Sitter supergravity \cite{Pilch:1984aw}: in both cases, while classical supersymmetric formulations exist \cite{Pilch:1984aw,Boulanger:2026wnw,Sezgin:2012ag,Gelfond:2026hrq}, the reality conditions needed for bulk fermions are inconsistent with a positive norm for the odd-spin bosons. 
	\item Can dS/CFT contain odd integer spins? In HS Gravity, the answer turns out to be ``yes'': without fermions, there exists a choice of reality conditions that allows the inclusion of odd spins, in the same HS-multiplet as the even spins, with positive norm. The bulk version of these boundary conditions was given in \cite{Chang:2013afa}, and the boundary version will be presented here. Thus, HS Gravity provides a loophole out of the no-go situation \cite{Pilch:1984aw} in supergravity, where spin-1 and spin-2 can share a multiplet only through fermions and supersymmetry. After the first version of our manuscript came out, a similar result was obtained in \cite{Gelfond:2026hrq}, where classical supersymmetric formulations are given whose bosonic sectors (both even and odd spins) are unitary.
	\item Can we reconcile the Euclidean-based holographic prescription of \cite{Maldacena:2002vr}, where only one of the two boundary data for each field is independent, with de Sitter QFT, where both are independent? Almost equivalently, can we reconcile the single boundary of Euclidean AdS with the two boundaries of de Sitter? In HS Gravity, we will make concrete progress on this question, by casting the type-A and type-B theories as simultaneously independent (in de Sitter) and equivalent (in Euclidean AdS). This will involve three main ingredients:
	\begin{enumerate}
		\item Extending the dS/CFT of \cite{Anninos:2011ui} to type-B.
		\item Using antipodal identifications of de Sitter space \cite{Parikh:2002py,Neiman:2014npa,Halpern:2015zia}.
		\item Invoking the ``dark magic'' of the HS-algebraic partition function at finite sources \cite{David:2020ptn}, which will yield the otherwise-unexpected equivalence between type-A and type-B.
	\end{enumerate}
\end{enumerate}
We were motivated to tackle these questions in Higher-Spin dS/CFT by some curious results in our previous work \cite{Lang:2024dkt} on the AdS/CFT correlators in the $\calN=2$ theory. In \cite{Lang:2024dkt}, we speculated on the potential relevance of those results for de Sitter. The present paper is essentially a matured version of those speculations.

Another question addressed in this paper is more intrinsic to HS theory: 
\begin{enumerate}[resume] 
	\item In de Sitter space, how is the unitary/Hermitian structure of the HS particle multiplet related to HS algebra?
\end{enumerate}
In AdS, the answer is well-known \cite{Konshtein:1988yg}, and consists of two steps. First, one constructs the Hilbert spaces of \emph{boundary} particles (bosons and/or fermions) as unitary representations of HS algebra. Then, via the Flato-Fronsdal theorem \cite{Flato:1978qz}, one constructs the bulk HS multiplet as the product of two such boundary representations. Now, in de Sitter, there is no Flato-Fronsdal theorem. Indeed, since the boundary is spacelike, there is no Hilbert space of boundary particles at all -- only a Euclidean CFT \footnote{However, see \cite{Anninos:2026hia} for a path-integral relation between the particles on a Euclidean $S_3$ boundary and unitary representations on Lorentzian $dS_3$. Leaving the fundamental boundary particles aside, one can try and associate a Hilbert space to the vector-model CFT itself, as if it were the analytic continuation of an ordinary Lorentzian CFT \cite{Ng:2012xp}. Note also the independent attempt \cite{Anninos:2017eib} to define a Hilbert space for the bulk HS gravity by re-imagining the CFT's ingredients and their roles.}. Instead, as we'll show, the bulk Hilbert space can be constructed directly within HS algebra, with a simple, manifestly HS-symmetric formula for the Hermitian norm. In fact, the complex-conjugation structure that makes this norm positive-definite is the same one that will allow us to include odd spins with positive norm into dS/CFT. Our positive-definite single-particle Hilbert space can include fermions without any issue, and obeys a (slightly exotic) supersymmetry. The problem with fermions and supersymmetry arises when we upgrade from particle wavefunctions to fields: whereas the former are complex, the latter require reality conditions, and as already mentioned, the ones consistent with fermions are not the ones that lead to positive-definite norm.

\subsection{Comparison with previous works on extending Higher-Spin dS/CFT}

Before proceeding, we must mention two previous works \cite{Chang:2013afa,Hertog:2017ymy}. These claimed to extend Higher-Spin dS/CFT from minimal type-A to the entire range of bosonic theories \cite{Chang:2013afa}, and to the $\calN=2$ theory \cite{Hertog:2017ymy}, thus overlapping with Questions 1-2 in our list above. Let us perform a quick comparison, to show that the corresponding sections of the present paper are non-redundant and necessary:
\begin{itemize}
	\item In \cite{Chang:2013afa}, the correct reality conditions for the bulk fields were found, as we'll review in section \ref{sec:reality}. However, the correct reality conditions on the boundary (for odds spins and for type-B) were missed, due to a naive reliance on analytic continuation from Lorentzian AdS/CFT. For example, while it was noticed that the bulk reality conditions for odd spins should be non-standard, no corresponding adjustment was made to the boundary fields and currents. Thus, boundary theories with correct signs of the 2-point functions remain to be constructed.
	\item In \cite{Hertog:2017ymy}, the reality conditions needed to give all the bosons positive norm weren't explicitly given, neither in the bulk nor on the boundary (though the possibility of unusual reality conditions, and their necessity with regard to odd spins, are mentioned). As a result, it wasn't noticed that these reality conditions are incompatible with bulk fermions, so that only the bosonic sector is consistent.
	\item At finite sources, the authors of \cite{Hertog:2017ymy} work with the standard CFT partition function, which suffers from the absence of global maximum noticed in \cite{Anninos:2012ft} (the authors of \cite{Hertog:2017ymy} observe that the problem gets better for supersymmetry-preserving sources, but that doesn't change the situation for the original non-supersymmetric sources of \cite{Anninos:2012ft}). In the present paper, we use the partition function prescription of \cite{David:2020ptn}, which leads to a global maximum (at least for sources of the lowest spherical harmonics). 
\end{itemize}
We also note the recent paper \cite{Anninos:2026hia} that proposes a novel, exotic kind of dS/CFT involving Euclidean de Sitter (i.e. the 4-sphere). In that work, a supersymmetric version of the duality has also been considered, though the question of its unitarity hasn't been fully addressed.

\subsection{Structure of the paper}

The rest of the paper is structured as follows. In section \ref{sec:geometry}, we review and define our conventions for embedding-space geometry, twistors, and HS algebra. In section \ref{sec:reality}, we review the non-standard complex conjugation and reality conditions of \cite{Chang:2013afa} for HS algebra. We discuss how these reality conditions extend to the non-linear Vasiliev equations, but fail to be consistent with supersymmetry. In section \ref{sec:Hermitian}, we present the Hilbert-space structure with positive-definite norm for the single-particle HS (super)multiplet in de Sitter, using HS algebra and the complex conjugation from section \ref{sec:reality}. In section \ref{sec:boundary}, we use this structure to motivate new reality conditions for the boundary vector model, which produce 2-point functions of the correct sign for all the integer-spin fields of the type-A and type-B theories in de Sitter. Using the techniques of \cite{Neiman:2017mel,David:2020ptn,Lang:2024dkt}, we construct the $n$-point correlators of these dS/CFT boundary theories using HS-algebra products of boundary-bulk (or boundary-twistor) propagators. In section \ref{sec:EAdS}, we switch from correlators to the partition function at finite sources, using the HS-algebraic prescription of \cite{David:2020ptn}. In section \ref{sec:EAdS:spin_1}, we show that the global-maximum property found for the (lowest spherical harmonic of) the spin-0 source in \cite{David:2020ptn} extends to the spin-1 source. In section \ref{sec:EAdS:Legendre}, we point out the Legendre-self-duality of the HS partition function, and describe how the type-A and type-B theories in Euclidean AdS can be viewed as the same theory with opposite boundary conditions. In section \ref{sec:elliptic_dS}, we present the complementary point of view in de Sitter space, where the type-A and type-B theories (each with its preferred boundary conditions) live on two ``parallel'' quotient spaces under the antipodal map. Section \ref{sec:discuss} is devoted to discussion and outlook.

\section{Embedding space, HS algebra, and Penrose transform} \label{sec:geometry}

In this section, we review the approach of \cite{Neiman:2017mel} to embedding-space geometry, twistors, and HS algebra. We closely follow the conventions of \cite{Neiman:2017mel,David:2020ptn}, but with factors of $i$ on the embedding-space coordinates in some places, so as to make Lorentzian de Sitter (rather then Euclidean AdS) the real spacetime. In that respect, our conventions are closer to those in \cite{Neiman:2013hca,Neiman:2018ufb}. 

\subsection{Spacetime}

We consider Lorentzian de Sitter spacetime $dS_4$, whose asymptotic boundary is a pair of conformal 3-spheres $\scri^\pm$. We represent these as embedded in a flat $\bbR^{1,4}$ Minkowski space, with a (mostly-plus) flat metric $\eta_{\mu\nu}$, which is used to raise and lower spacetime indices $(\mu,\nu,\dots)$. The $dS_4$ bulk is defined as the hyperboloid of unit spacelike vectors in $\bbR^{1,4}$:
\begin{align}
	dS_4 = \left\{x^\mu\in\bbR^{1,4}\, |\, x_\mu x^\mu = 1 \right\} \ . \label{eq:dS}
\end{align}
Vectors in $dS_4$ are simply vectors $v^\mu\in \bbR^{1,4}$ that are tangential to the hyperboloid \eqref{eq:dS}, i.e. that satisfy $v\cdot x \equiv v_\mu x^\mu = 0$. The covariant derivative $\nabla_\mu$ on $dS_4$ is just the flat derivative $\del_\mu$ in $\bbR^{1,4}$, followed by projecting all tensor indices back into the $dS_4$ tangent space with the projector $q_\mu^\nu(x) = \delta_\mu^\nu - x_\mu x^\nu$.

The future (past) conformal boundary of $dS_4$ is the 3-sphere of future-pointing (past-pointing) lightlike directions in $\bbR^{1,4}$, i.e. the lightcone modulo rescalings:
\begin{align}
	\scri^\pm = \left. \left\{\ell^\mu\in \bbR^{1,4}\, |\, \ell_\mu \ell^\mu = 0, \ \ell^0 \gtrless 0 \right\} \middle/ \left\{\ell^\mu \to \rho\ell^\mu \right\} \right. \ . \label{eq:scri}
\end{align}
The rescalings $\ell^\mu \rightarrow \rho\ell^\mu$ induce Weyl transformations $d\ell_\mu d\ell^\mu\rightarrow \rho^2 d\ell_\mu d\ell^\mu$ on the boundary's conformally flat metric. A boundary quantity that depends on the scaling as $f(\rho\ell^\mu) = \rho^{-\Delta}f(\ell^\mu)$ is said to have conformal weight $\Delta$. Boundary \emph{vectors} at a point $\ell$ can be described as embedding-space vectors $v^\mu\in \bbR^{1,4}$ that are tangential to the lightcone, i.e. $v\cdot\ell = 0$, modulo the equivalence relation $v^\mu\cong v^\mu + \alpha\ell^\mu$. 

A conformal frame on the boundary corresponds to choosing a particular vector $\ell^\mu$ along each null direction, i.e. choosing a section of the $\bbR^{1,4}$ lightcone. Two such conformal frames will be especially useful. One is defined by the section $\ell^0 = \pm 1$, which gives $\scri^\pm$ the metric of a unit 3-sphere $S_3$. The other is a flat $\bbR^3$ frame, defined for $\scri^\pm$ as:
\begin{align}
	\ell^\mu(\mathbf{r}) = \pm \left(\frac{1+\mathbf{r}^2}{2}, \frac{1-\mathbf{r}^2}{2}, \mathbf{r} \right) \ . \label{eq:Poincare}
\end{align}
In addition to the real de Sitter spacetime \eqref{eq:dS}, we will make use of its analytic continuation into imaginary embedding-space coordinates $x^\mu$. This forms a pair of Euclidean AdS spaces, one at negative and one at positive imaginary time:
\begin{align}
	EAdS_4^\pm = \left\{x^\mu\in i\bbR^{1,4}\, |\, x_\mu x^\mu = 1, \ \Im x^0 \lessgtr 0 \right\} \ . \label{eq:EAdS}
\end{align} 
Each of the two spaces $EAdS_4^\pm$ has a single conformal 3-sphere as its boundary. In embedding space, these 3-spheres are just the negative/positive imaginary multiples of e.g. $\scri^+$. Since \eqref{eq:EAdS} is just an analytic continuation of \eqref{eq:dS}, all the bulk equations below that don't involve reality or complex conjugation will be equally valid in de Sitter and in Euclidean AdS. The significance of the Euclidean AdS spaces \eqref{eq:EAdS} is that positive-frequency fields in $dS_4$ have singularities when continued to $EAdS_4^-$ but not when continued to $EAdS_4^+$, and vice versa for negative frequencies. 

\subsection{Twistor space} \label{sec:geometry:twistors}

The isometry group of $dS_4$ and $EAdS_4$, or the conformal group of their boundaries, is just the rotation group $SO(1,4)$ in the embedding space $\bbR^{1,4}$. The double cover of this group is $USp(2,2)$. Its spin-$\frac{1}{2}$ representation, composed of 4-component Dirac spinors, is what we'll refer to as \emph{twistor space} $\bbT$. We will denote twistors by the indices $(a,b,\dots)$. Twistor space has a symplectic metric $I_{ab}$ with inverse $I^{ab}I_{ac} = \delta^b_c$, which we use to raise and lower indices as $U_a = I_{ab}U^b$, $U^a = U_b I^{ba}$. From $I_{ab}$, we can construct a Levi-Civita symbol $\epsilon_{abcd} = 3I_{[ab}I_{cd]}$, along with a measure $d^4U$ (in which we include $2\pi$ factors for convenience):
\begin{align}
	d^4U \equiv \frac{\epsilon_{abcd}}{4!(2\pi)^2}\,dU^a dU^b dU^c dU^d \ . 
\end{align} 
Twistor space and spacetime are related by the gamma matrices $(\gamma_\mu)^a{}_b$, which satisfy the Clifford algebra $\{\gamma_\mu,\gamma_\nu\} = -2\eta_{\mu\nu}$. We also introduce the antisymmetrized products $\gamma_{\mu\nu} \equiv \frac{1}{2}[\gamma_\mu,\gamma_\nu]$, which generate the $SO(1,4)$ spacetime symmetry. As twistor matrices, the $\gamma_\mu$ are traceless and antisymmetric, while the $\gamma_{\mu\nu}$ are symmetric. We use $\gamma_\mu$ to map between twistor matrices and $\bbR^{1,4}$ vectors, and similarly for $\gamma_{\mu\nu}$ and $\bbR^{1,4}$ bivectors:
\begin{align}
	\xi^{ab} = \gamma_\mu^{ab}\xi^\mu \ ; \quad \xi^\mu = -\frac{1}{4}\gamma^\mu_{ab}\xi^{ab} \ ; \quad 
	f^{ab} = \frac{1}{2}\gamma_{\mu\nu}^{ab}f^{\mu\nu} \ ; \quad f^{\mu\nu} = \frac{1}{4}\gamma^{\mu\nu}_{ab} f^{ab} \ .
\end{align}
In our $SO(1,4)$ signature, the properties of twistors under complex conjugation follow those of $SO(3)$ spinors. The complex conjugation operation $U^a\rightarrow \bar U^a$ that is consistent with $SO(1,4)$ symmetry squares to $-1$:
\begin{align}
	\bar{\bar U}^a = -U^a \ .
\end{align}
Therefore, it can't be identified with \emph{component-wise} complex conjugation $\overline{U^a}$, and there is no invariant notion of a real twistor. Explicitly, we can set:
\begin{align}
	\bar U_a &= \begin{pmatrix} 1 & 0 \\ 0 & -1 \end{pmatrix} \overline{U^b} \ , \label{eq:U_bar}
\end{align}
where we use $2\times 2$ block notation, so that ``$1$'' denotes a $2\times 2$ identity matrix. The symplectic metric $I_{ab}$ can be defined as real, both component-wise and under the twistor complex conjugation, $\bar I_{ab} = I_{ab}$. The gamma matrices $(\gamma_\mu)^a{}_b$ are real under the twistor complex conjugation $\bar\gamma_\mu = \gamma_\mu$, but can't all be real component-wise. Instead, they are quaternionic, i.e. they can be written in $2\times 2$ blocks made up of $(1,-i\boldsymbol{\sigma})$, where $\boldsymbol{\sigma}$ are the Pauli matrices. In particular, we can set:
\begin{align}
		I_{ab} &= \begin{pmatrix} -i\sigma_2 & 0 \\ 0 & i\sigma_2 \end{pmatrix} \ ; \quad
		(\gamma_\mu)^a{}_b = \left( \begin{pmatrix} 1 & 0 \\ 0 & -1 \end{pmatrix} , \begin{pmatrix} 0 & -1 \\ 1 & 0 \end{pmatrix} , \begin{pmatrix} 0 & -i\boldsymbol{\sigma} \\ -i\boldsymbol{\sigma} & 0 \end{pmatrix} \right) \ . \label{eq:I_gamma}
\end{align}
Since $I_{ab}$ is antisymmetric, the inner product of any twistor with itself vanishes $U_a U^a = 0$. In contrast, the real norm $\bar U_a U^a$ is generally non-vanishing, and has signature $(+,+,-,-)$, as manifest in \eqref{eq:U_bar}. The subsets of $\bbT$ with positive/negative/vanishing norm $\bar U_a U^a$ are referred to as $\bbT^+$, $\bbT^-$ and $\bbN$, respectively. 

We will often use condensed index-free notation, in which twistor indices are implicitly contracted bottom-to-top. Thus, for example, $Ux\ell V \equiv U_a(x^\mu\gamma_\mu)^a{}_b(\ell^\nu\gamma_\nu)^b{}_c V^c$. We define the delta function $\delta(U)$ on twistor space via the integral:
\begin{align}
	\delta(U) = \int d^4V e^{iVU} \ ; \quad \int d^4U\,\delta(U) f(U) = f(0) \ . \label{eq:delta_U}
\end{align}
Note that integrals such as \eqref{eq:delta_U} are formal, since the twistor space $\bbT$ is complex, and there is no preferred contour for the integration.

\subsection{Local spinor spaces} \label{sec:geometry:spinors}

Choices of (bulk or boundary) spacetime points pick out various $\bbC^2$ subspaces out of the twistor space $\bbT=\bbC^4$. These are the local spinor spaces at each spacetime point. In fact, Penrose essentially \emph{defines} spacetime points as $\bbC^2$ subspaces of twistor space (though these are usually described as $\bbC\bbP^1$, since he prefers to work projectively). To see how this works, consider a spacetime point represented by a vector $\xi^\mu \in \bbR^{1,4}$. The $\bbC^2$ subspace associated with the point $\xi^\mu$ is spanned by the rank-2 twistor matrix:
\begin{align}
	P^a{}_b(\xi) = \frac{1}{2}\left(\sqrt{\xi\cdot\xi}\,\delta^a_b + i\xi^a{}_b\right) \ ,
\end{align}
or $P(\xi) = \frac{1}{2}\left(\sqrt{\xi\cdot\xi} + i\xi \right)$ in index-free notation. We also use the notation $P(\xi)$ for the $\bbC^2$ subspace itself. When we want to emphasize that a twistor belongs to the subspace $P(\xi)$, we denote it as e.g. $u^a_{(\xi)}$ rather than simply $U^a$. The subspaces $P(\xi)$ and $P(-\xi)$ are totally orthogonal under the twistor metric. 

A measure $d^2u_{(\xi)}$ on $P(\xi)$ can be defined as:
\begin{align}
	\frac{du_{(\xi)}^a du_{(\xi)}^b}{2\pi} \equiv P^{ab}(\xi)\, d^2u_{(\xi)} \ . \label{eq:measure}
\end{align}
With this measure, we define a delta function, whose support is on $P(-\xi)$, i.e. on the subspace orthogonal to $P(\xi)$:
\begin{align}
	\delta_\xi(U) = \int_{P(\xi)} d^2v_{(\xi)}\,e^{iv_{(\xi)}U} \ . \label{eq:delta_xi}
\end{align}
The HS-algebra products that will be relevant for correlators and partition functions below all follow from two spinor-integral formulas. The first is the integral of a delta function associated with one spacetime point $\xi$ over the spinor space associated with another point $\xi'$:
\begin{align}
	\int_{P(\xi')} d^2u_{(\xi')}\,\delta_{\xi}(u_{(\xi')}) f(u_{(\xi')}) = \frac{2}{\sqrt{(\xi\cdot\xi)(\xi'\cdot\xi')} + \xi\cdot\xi'}\,f(0) \ . \label{eq:spinor_delta_integral}
\end{align}
The second is the Gaussian integral on $P(\xi)$:
\begin{align}
	\int_{P(\xi)} d^2u_{(\xi)}\, e^{u_{(\xi)}Au_{(\xi)}/2} = \frac{\pm 1}{\sqrt{\det_\xi(A)}} \ ; \quad \det\nolimits_\xi(A) = -\frac{1}{2}\tr\left(P(\xi)A\right)^2 \ , \label{eq:Gaussian_spinor}
\end{align}
where $A_{ab}$ is a symmetric twistor matrix. The sign ambiguity in complex Gaussian integrals \eqref{eq:Gaussian_spinor} plays a key role \cite{David:2020ptn} in the difference between the standard CFT partition function and the HS-algebraic one that we will use in section \ref{sec:EAdS}. 

Let's now consider more concretely the local spinor spaces at bulk vs. boundary points. At a bulk point, we take $\xi^\mu = x^\mu$ with $x\cdot x = 1$. The choice of point $x$ breaks the spacetime symmetry $SO(1,4)$ down to the bulk Lorentz group $SO(1,3)$ at $x$. Accordingly, twistor space decomposes into the right-handed and left-handed Weyl spinor spaces at $x$. The rank-2 twistor matrices $P(\pm x) = \frac{1}{2}(1\pm ix)$ are projectors, which span (or project onto) the right-handed and left-handed spinor space. The restriction of the twistor metric $I_{ab}$ onto these spinor spaces is simply $P_{ab}(\pm x)$. 

The $dS_4$ covariant derivative $\nabla_\mu$ can now be extended to objects with Weyl spinor indices: it is again the flat derivative $\del_\mu$, followed by projecting every tensor index back into the tangent space at $x$ using $q_\mu^\nu(x) = \delta_\mu^\nu - x_\mu x^\nu$, and every spinor index back into the appropriate spinor space using $P(\pm x) = \frac{1}{2}(1\pm ix)$. It is convenient to define a covariant derivative with spinor indices as:
\begin{align}
	\hat\nabla_{ab} \equiv P^c{}_a(x) P^d{}_b(-x) \gamma^\mu_{cd} \nabla_\mu \ ,
\end{align}
so that the right-handed spinor index is forced into the first position, and the left-handed one into the second position.

Now, consider a choice of \emph{boundary} point, i.e. $\xi^\mu = \ell^\mu$ with $\ell\cdot\ell = 0$. This breaks $SO(1,4)$ down to $ISO(3)$. Twistor space now acquires only \emph{one} preferred $\bbC^2$ subspace -- the space spanned by the matrix $P(\ell) = \frac{i}{2}\ell$, which scales with $\ell^\mu$. The subspace spanned by $P(\ell)$ is totally null under the twistor metric $I_{ab}$. From the point of view of the 3d boundary, $P(\ell)$ is the space of \emph{cospinors} at $\ell$; in particular, the measure \eqref{eq:measure} on it scales inversely with $\ell^\mu$. 

Under the twistor complex conjugation, we have $\bar P(\xi) = P(-\bar\xi)$, where $\xi^\mu\to\bar\xi^\mu$ is the standard complex conjugation of vectors. Thus, for real bulk points $x^\mu\in dS_4$ we have $\bar P(x) = P(-x)$, reflecting the complex-conjugate relation between left-handed and right-handed spinors of $SO(1,3)$; for boundary points $\ell^\mu\in\scri^\pm$, we have $\bar P(\ell) = P(-\ell) = -P(\ell)$, reflecting the fact that the spinor space of $SO(3)$ is closed under complex conjugation; and for Euclidean bulk points $x^\mu\in EAdS_4^\pm$, we have $\bar P(x) = P(x)$, reflecting the closure under complex conjugation of the left-handed and right-handed spinor spaces of $SO(4)$. 

In terms of the real twistor norm $\bar U_a U^a$, the subspaces spanned by $P(x)$ with $x\in dS_4$, or by $P(\ell)$ with $\ell\in \scri^\pm$, all lie within the null subset $\bbN$. All nonzero elements of the subspaces spanned by $P(x)$ with $x\in EAdS^\pm$ lie in the subset $\bbT^\pm$, respectively.

\subsection{HS algebra} \label{sec:correlators:HS:algebra}

Consider again the Clifford algebra $\{\gamma_\mu,\gamma_\nu\} = -2\eta_{\mu\nu}$. Higher-spin algebra is just an application of the same concept, but to a twistor $Y^a$ instead of a vector $\gamma^\mu$. Since the twistor metric is antisymmetric, we get a commutator $[Y_a,Y_b]_\star = 2iI_{ab}$ instead of an anti-commutator. Imposing a symmetric ordering convention, we arrive at the following non-commutative product:
\begin{align}
	Y^a\star Y^b = Y^a Y^b + iI^{ab} \ . \label{eq:star}
\end{align}
This can be extended to polynomials as:
\begin{align}
	f(Y)\star g(Y) = f \exp\left(iI^{ab}\overleftarrow{\frac{\del}{\del Y^a}}\overrightarrow{\frac{\del}{\del Y^b}}\right) g \ . \label{eq:star_diff}
\end{align}
and to more general functions (and distributions) as:
\begin{align}
	f(Y)\star g(Y) = \int d^4U d^4V f(Y+U)\, g(Y+V)\, e^{-iUV} \ . \label{eq:star_int}
\end{align}
The bosonic HS symmetry algebra is the infinite-dimensional Lie algebra of even (i.e. integer-spin) functions $f(Y)$ with the associative product \eqref{eq:star}-\eqref{eq:star_int}. The quadratic elements $Y_a Y_b$ generate the spacetime symmetry group $SO(1,4)$, in analogy to the role of $\gamma_{[\mu}\gamma_{\nu]}$ in Clifford algebra. In supersymmetric versions of HS algebra, one allows also odd functions $f(Y)$. Usually, these are paired with extra Grassmann coordinates (or, in Vasiliev's original version, external Klein operators). However, as we saw for $\calN=2$ correlators in \cite{Lang:2024dkt}, and will see again for a $\calN=1$ Hilbert space in section \ref{sec:Hermitian}, for some purposes one can handle supersymmetry without such extra ingredients, using only the algebra \eqref{eq:star}-\eqref{eq:star_int}.

Another important element of HS algebra is the delta function $\delta(Y)$, as defined in \eqref{eq:delta_U}. A star product with $\delta(Y)$ implements a Fourier transform in twistor space:
\begin{align}
	f(Y)\star\delta(Y) = \int d^4U f(U)\,e^{iUY} \ ; \quad \delta(Y)\star f(Y) = \int d^4U f(U)\,e^{-iUY} \ . \label{eq:delta_Fourier}
\end{align}
$\delta(Y)$ squares to unity $\delta(Y)\star\delta(Y) = 1$, and (anti)commutes with even (odd) functions $f(Y)$, as:
\begin{align}
	\delta(Y)\star f(Y)\star\delta(Y) = f(-Y) \ . \label{eq:delta_flip}
\end{align}
The product \eqref{eq:star}-\eqref{eq:star_int} respects a supertrace operation, defined simply by:
\begin{align}
	\tr_\star f(Y) = f(0) \ . \label{eq:str}
\end{align}
For even functions, the operation \eqref{eq:str} satisfies the cyclic trace property $\tr_\star(f\star g) = \tr_\star(g\star f)$. When odd functions are included, this becomes the supertrace property:
\begin{align}
	\tr_\star\big(f(Y)\star g(Y)\big) = \tr_\star\big(g(-Y)\star f(Y)\big) \ .
\end{align}
Now, recall that choices of spacetime points, i.e. of embedding-space vectors $\xi^\mu\in\bbR^{1,4}$, pick out some subspaces of twistor space, which describe local spinors. The delta functions $\delta_\xi(Y)$ associated with these spinor spaces have the following properties under the HS star product:
\begin{align}
	&f(Y)\star\delta_\xi(Y) = \int_{P(\xi)} d^2u_{(\xi)}\,f(Y+u_{(\xi)})\,e^{iu_{(\xi)}Y} \ ; \label{eq:delta_Fourier_xi_first} \\ 
	&\delta_\xi(Y)\star f(Y) = \int_{P(\xi)} d^2u_{(\xi)}\,f(Y+u_{(\xi)})\,e^{-iu_{(\xi)}Y} \ ; \label{eq:delta_Fourier_xi_second} \\
	&\delta_\xi(Y)\star\delta(Y) = \delta(Y)\star\delta_\xi(Y) = \delta_{-\xi}(Y) \ . \label{eq:spinor_twistor_delta}
\end{align}
The star product of a pair of spinor delta functions reads:
\begin{align}
	\delta_\xi(Y)\star\delta_{\xi'}(Y) = \frac{2}{\sqrt{(\xi\cdot\xi)(\xi'\cdot\xi')} + \xi\cdot\xi'}\,\exp\left(\frac{iY\xi\xi'Y/2}{\sqrt{(\xi\cdot\xi)(\xi'\cdot\xi')} + \xi\cdot\xi'}\right) \ . \label{eq:general_two_point}
\end{align}
Star products with any additional delta functions will continue to result in Gaussians. A particularly simple case is when all points are on the boundary, i.e. when the vectors $(\xi^\mu,\xi'^\mu,\dots)$ are null $(\ell^\mu,\ell'^\mu,\dots)$. Here, the 3-point product reduces to the 2-point product:
\begin{align}
	\delta_\ell(Y)\star\delta_{\ell'}(Y) &= \frac{2}{\ell\cdot\ell'}\exp\frac{iY\ell\ell' Y}{2\ell\cdot\ell'} \ ; \label{eq:2_pt_product} \\
	\delta_\ell(Y)\star\delta_{\ell'}(Y)\star\delta_{\ell''}(Y) &= \pm\sqrt{-\frac{\ell\cdot\ell''}{2(\ell\cdot\ell')(\ell'\cdot\ell'')}} \, \delta_\ell(Y)\star\delta_{\ell''}(Y) \ . \label{eq:3_pt_product}
\end{align}
The sign ambiguity in \eqref{eq:3_pt_product}, which results from a Gaussian integral of the form \eqref{eq:Gaussian_spinor}, will be important for the distinction \cite{David:2020ptn} between the CFT partition function and the HS-algebraic one. The 3-point product \eqref{eq:3_pt_product} is a special case of the ``forgetful property'':
\begin{align}
	\delta_\ell(Y)\star f(Y)\star\delta_{\ell'}(Y) = \frac{\ell\cdot\ell'}{2}\tr_\star\big(\delta_\ell\star f\star\delta_{\ell'}\big)\,\delta_\ell(Y)\star\delta_{\ell'}(Y) \ .
\end{align}
We now turn to delta functions $\delta_x(Y)$ associated with a bulk point $x^\mu$. These satisfy:
\begin{align}
	\delta_x(Y)\star\delta_x(Y) = 1 \ ; \quad \delta_x(Y)\star\delta_{-x}(Y) = \delta(Y) \ . \label{eq:delta_delta}
\end{align}
The star-products \eqref{eq:delta_Fourier_xi_first}-\eqref{eq:delta_Fourier_xi_second} at a bulk point $x$ can be simplified by shifting the integration variable. Specifically, if we decompose the twistor argument as $Y = y_{(x)} + y_{(-x)}$ with $y_{(x)}\in P(x)$ a right-handed spinor at $x$ and $y_{(-x)}\in P(-x)$ a left-handed one, then eqs. \eqref{eq:delta_Fourier_xi_first}-\eqref{eq:delta_Fourier_xi_second} simply Fourier-transform $y_{(x)}$ while leaving $y_{(-x)}$ untouched:
\begin{align}
	f(Y)\star\delta_x(Y) & = \int_{P(x)} d^2u_{(x)}\,f(u_{(x)}+y_{(-x)})\,e^{iu_{(x)} y_{(x)}} \ ; \label{eq:bulk_Fourier_first} \\ 
	\delta_x(Y)\star f(Y) &= \int_{P(x)} d^2u_{(x)}\,f(u_{(x)}+y_{(-x)})\,e^{-iu_{(x)} y_{(x)}} \label{eq:bulk_Fourier_second} \ .
\end{align}
Combining these, we see that conjugation by $\delta_x(Y)$ flips the sign of $y_{(x)}$, similar to \eqref{eq:delta_flip}:
\begin{align}
	\delta_x(Y)\star f(Y)\star\delta_x(Y) = f(-ixY) = f(-y_{(x)} + y_{(-x)}) \ . \label{eq:bulk_delta_flip}
\end{align}

\subsection{The Penrose transform} \label{sec:geometry:Penrose}

The Penrose transform relates twistor functions $F(Y)$ to solutions of the free massless field equations (of all spins) in the $dS_4$ bulk. In HS language, the Penrose transform is just a star-product of the form \eqref{eq:delta_Fourier_xi_first}:
\begin{align}
	C(x;Y) = iF(Y)\star\delta_x(Y) \ , \label{eq:Penrose}
\end{align}
where we recall that $\delta_x(Y)$ is the delta function on the right-handed Weyl spinor space at $x$, and the star-product in \eqref{eq:Penrose} acts as a Fourier transform \eqref{eq:bulk_Fourier_first} on the right-handed spinor argument $y_{(x)} = P(x)Y$.

The twistor function $F(Y)$ in \eqref{eq:Penrose} has no spacetime dependence whatsoever. Through the transform \eqref{eq:Penrose}, it is converted into an $x$-dependent \emph{master field} $C(x;Y)$, which encodes the higher-spin field strengths at $x$, as well as their derivatives. The factor of $i$ follows the conventions of \cite{Neiman:2017mel}. The transform \eqref{eq:Penrose} is easy to invert, thanks to the identity \eqref{eq:delta_delta}. The fields of different helicities $h$ enter the twistor function $F(Y)$ as different degrees of homogeneity $-2h-2$ w.r.t. the twistor argument:
\begin{align}
	Y^a\frac{\del F(Y)}{\del Y^a} = (-2h-2)F(Y) \ . \label{eq:homogeneity}
\end{align}
Let's now describe how the various bulk master fields are encoded inside $C(x;Y)$. The right-handed spin-$s$ field strength at $x$ is an object $C^{(s)}_{a_1\dots a_{2s}}(x)$ with $2s$ totally symmetrized spinor indices, all of which lie in the right-handed spinor space $P(x)$. Similarly, the left-handed field strength $C^{(-s)}_{a_1\dots a_{2s}}(x)$ has $2s$ totally symmetrized indices in the left-handed spinor space $P(-x)$. For spin 0, there is no right-handed/left-handed distinction, and we have just the scalar field $C^{(0)}(x)$. To extract these field strengths from the master field $C(x;Y)$, we first decompose the twistor argument $Y=y_{(x)}+y_{(-x)}$ into its Weyl-spinor pieces $y_{(x)}\in P(x)$ and $y_{(-x)}\in P(-x)$. The field strengths are now encoded as the $2s$'th powers of $y_{(x)}$ and $y_{(-x)}$, respectively:
\begin{align}
	\begin{split}
		C(x;y_{(x)}) &= \sum_{s=0}^\infty \frac{1}{(2s)!}\,C^{(s)}_{a_1\dots a_{2s}}(x)\, y_{(x)}^{a_1}\dots y_{(x)}^{a_{2s}} \ ; \\
		C(x;y_{(-x)}) &= \sum_{s=0}^\infty \frac{1}{(2s)!}\,C^{(-s)}_{a_1\dots a_{2s}}(x)\, y_{(-x)}^{a_1}\dots y_{(-x)}^{a_{2s}} \ .
	\end{split} \label{eq:packaging}
\end{align}
The right-handed and left-handed gauge field strengths \eqref{eq:packaging} can be combined into a field-strength tensor for each spin, via:
\begin{align}
	C^{(s)}_{\mu_1\nu_1\cdots\mu_s\nu_s}(x) = \frac{1}{4^s}\,\gamma_{\mu_1\nu_1}^{a_1 b_1}\dots \gamma_{\mu_s\nu_s}^{a_s b_s}\left( C^{(s)}_{a_1 b_1\dots a_s b_s}(x) + C^{(-s)}_{a_1 b_1\dots a_s b_s}(x) \right) \ ,
	\label{eq:Weyl_tensor}
\end{align}
while $C^{(0)}(x) = C(x;0)$ is the spin-0 field. Eq. \eqref{eq:Weyl_tensor} describes a Maxwell field strength for $s=1$, a linearized Weyl tensor for $s=2$, and their generalizations for higher spins. The Penrose transform \eqref{eq:Penrose} automatically ensures that these fields satisfy the free massless field equations in $dS_4$:
\begin{align}
	\begin{split}
		s=0: \quad &\nabla_\mu\nabla^\mu C^{(0)}(x) = 2C^{(0)}(x) \ ; \\
		s=1: \quad &\nabla^\mu C^{(1)}_{\mu\nu}(x) = \nabla_{[\mu} C^{(1)}_{\nu\rho]}(x) = 0 \ ; \\
		s\geq 2: \quad &\nabla^{\mu_1} C^{(s)}_{\mu_1\nu_1\cdots\mu_s\nu_s}(x) = 0 \ .
	\end{split}
\end{align}
Note that the field equation for $C^{(0)}(x)$ is that of a \emph{conformally} massless scalar. 

Finally, the Taylor coefficients of $C(x;Y)$ that don't appear in \eqref{eq:packaging} are identified via the Penrose transform as derivatives of the fields \eqref{eq:packaging}:
\begin{align}
	\begin{split}
		\left.\frac{\del^{2(s+k)} C(x;Y)}{\del y_{(x)}^{a_1}\dots\del y_{(x)}^{a_{2s+k}}\del y^{(-x)}_{b_1}\dots\del y^{(-x)}_{b_k}}\right|_{Y=0} 
		&= \hat\nabla_{(a_1}{}^{(b_1}\dots\hat\nabla_{a_k}{}^{b_k)} C^{(s)}_{a_{k+1}\dots a_{2s+k})}(x) \ ; \\
		\left.\frac{\del^{2(s+k)} C(x;Y)}{\del y_{(x)}^{a_1}\dots\del y_{(x)}^{a_k}\del y^{(-x)}_{b_1}\dots\del y^{(-x)}_{b_{2s+k}}}\right|_{Y=0} 
		&= \hat\nabla_{(a_1}{}^{(b_1}\dots\hat\nabla_{a_k)}{}^{b_k} C_{(-s)}^{\,b_{k+1}\dots b_{2s+k})}(x) \ .
	\end{split} \label{eq:unfolding}
\end{align}
Knowing the master field at one point, we can calculate it at any other, using either the field equations or the Penrose transform \eqref{eq:Penrose}. A particularly simple case is the master field at the \emph{antipodal} point $-x^\mu$, which takes the form:
\begin{align}
	C(-x;Y) = C(x;Y)\star\delta(Y) \ , \label{eq:C_antipodal}
\end{align}
thanks to the star-product identity \eqref{eq:spinor_twistor_delta}. This implies that the antipodal map $x^\mu\to -x^\mu$ acts on the twistor function $F(Y)$ as:
\begin{align}
	F(Y)\to F(Y)\star\delta(Y) \ . \label{eq:F_antipodal}
\end{align}

\section{Complex conjugation and reality conditions for twistor functions and bulk fields} \label{sec:reality}

In this section, we discuss complex conjugation in HS algebra, and the corresponding reality conditions for fields in de Sitter. In section \ref{sec:reality:new}, after reviewing the standard complex conjugation \& reality conditions, we review the ``new'' ones introduced in \cite{Chang:2013afa}, which will lead to a positive-definite Hermitian norm for all spins (section \ref{sec:Hermitian}), and to dS/CFT 2-point functions of the correct sign for all integer spins (section \ref{sec:boundary}). In section \ref{sec:reality:Vasiliev}, we discuss how the ``new'' reality conditions of \cite{Chang:2013afa} extend to the non-linear bosonic Vasiliev equations. In section \ref{sec:reality:fermions}, we show that these reality conditions are \emph{not} consistent with fermions/supersymmetry. 

\subsection{Complex conjugation \& reality conditions, ``old'' and ``new''} \label{sec:reality:new}

The standard complex conjugation \cite{Iazeolla:2007wt,Sezgin:2012ag} considered for the HS algebra \eqref{eq:star}-\eqref{eq:star_int} in de Sitter is:
\begin{align}
	f^\dagger(Y) \equiv \overline{f(\bar Y)} \ . \label{eq:old_conjugation}
\end{align}
The reason for the ``$\dagger$ notation is that this conjugation reverses the order of the star-product:
\begin{align}
	(f\star g)^\dagger = g^\dagger\star f^\dagger \ .
\end{align}
Other basic identities read:
\begin{align}
	(f^\dagger)^\dagger(Y) = f(-Y) \ ; \quad \delta^\dagger(Y) = \delta(Y) \ ; \quad \delta_\xi^\dagger(Y) = \delta_{-\bar\xi}(Y) \ . \label{eq:dagger_identities}
\end{align}
Now, consider free massless fields in de Sitter, as in section \ref{sec:geometry:Penrose}. The standard reality condition for these is that the right-handed curvature is complex-conjugate to the left-handed one:
\begin{align}
	\bar C^{(s)}_{a_1\dots a_{2s}}(x) = \eta\,C^{(-s)}_{a_1\dots a_{2s}}(x) \ ; \quad \bar C^{(-s)}_{a_1\dots a_{2s}}(x) = (-1)^{2s}\,\eta\,C^{(s)}_{a_1\dots a_{2s}}(x) \ . \label{eq:reality_components}
\end{align}
Here, we introduced an overall sign factor $\eta = \pm 1$. Flipping $\eta\to-\eta$ interchanges the notions of real and imaginary fields. As we will see, it will be natural to set $\eta = +1$ for the type-A theory, and $\eta = -1$ for the type-B theory. This is in keeping with the reality properties of the type-A and type-B sectors in Vasiliev's original $\calN=2$ construction \cite{Vasiliev:1990en,Vasiliev:1992av}. 

At the level of the master field $C(x;Y)$, the reality conditions \eqref{eq:reality_components} can be expressed via the complex-conjugation operation \eqref{eq:old_conjugation}, with the help of \eqref{eq:bulk_delta_flip}, as:
\begin{align}
	C^\dagger(x;Y) = \eta\,\delta_x(Y)\star C(x;Y)\star\delta_x(Y) \qquad \text{(for a real point $x\in dS_4$)} \ . \label{eq:C_dagger}
\end{align}
In terms of the twistor function $F(Y)$, this implies:
\begin{align}
	F^\dagger(Y) = -\eta\,\delta(Y)\star F(Y) \ . \label{eq:F_dagger}
\end{align}

Now, crucially for the present paper, there is another kind of complex conjugation \cite{Chang:2013afa} that is respected by the HS algebra \eqref{eq:star}-\eqref{eq:star_int}:
\begin{align}
	f^*(Y) \equiv \overline{f(-i\bar Y)} \ . \label{eq:new_conjugation}
\end{align}
Unlike the standard conjugation \eqref{eq:old_conjugation}, the ``new'' conjugation \eqref{eq:new_conjugation} preserves the order of the star-product:
\begin{align}
	(f\star g)^* = f^*\star g^* \ . \label{eq:star_order}
\end{align}
The analogue of the identities \eqref{eq:dagger_identities} reads:
\begin{align}
	(f^*)^*(Y) = f(-Y) \ ; \quad \delta^*(Y) = \delta(Y) \ ; \quad \delta_\xi^*(Y) = -\delta_{-\bar\xi}(Y) \ . \label{eq:star_identities}
\end{align}
We can use the conjugation \eqref{eq:new_conjugation} to define an alternative reality condition on fields in de Sitter, in place of \eqref{eq:C_dagger}-\eqref{eq:F_dagger}:
\begin{align}
	C^*(x;Y) = \eta\,C(x;Y) \quad \text{(for a real point $x\in dS_4$)} \ ; \quad F^*(Y) = \eta\,F(Y)\star\delta(Y) \ . \label{eq:reality_star}
\end{align}
The conjugation \eqref{eq:new_conjugation} was briefly considered for the AdS theory (using slightly different conventions) in \cite{Chang:2012kt}, but was abandoned because the corresponding reality conditions do not allow for bulk fermions. As we will see in section \ref{sec:reality:fermions}, the same is true in de Sitter. For bosons with integer spin $s$, the ``new'' reality condition \eqref{eq:reality_star} reproduces the standard one \eqref{eq:reality_components}, but with a factor of $(-1)^s$. Thus, the reality condition for even spins remains the same, but for odd spins, the notions of real and imaginary fields are interchanged. 

Finally, let us note the Euclidean AdS version of the reality condition \eqref{eq:reality_star}. In Euclidean signature, the reality conditions on the bulk fields no longer relate the left-handed and right-handed field strengths as in \eqref{eq:reality_components}, but apply to each handedness separately. At the level of the master field $C(x;Y)$, the reality condition can still be written as $C^*(x;Y) = C(x;Y)$ (here, we will not need the overall sign factor $\eta$). However, since the coordinates $x^\mu$ are now imaginary $\bar x^\mu = -x^\mu$, the implication for the twistor function $F(Y)$ is different. Overall, the condition for real fields in Euclidean AdS (again, in a flipped sense for the case of odd spins) reads:
\begin{align}
	C^*(x;Y) = C(x;Y) \quad \text{(for an imaginary point $x\in EAdS_4$)} \ ; \quad F^*(Y) = F(Y) \ . \label{eq:reality_star_EAdS}
\end{align}

\subsection{``New'' reality condition vs. Vasiliev equations} \label{sec:reality:Vasiliev}

We now discuss how the ``new'' reality condition \eqref{eq:reality_star} on HS fields in $dS_4$ is consistent with the non-linear bosonic Vasiliev equations (or, slightly more generally, with the bosonic sector of the $\calN=2$ equations). For compatibility with the notations in the rest of this paper, we consider the embedding-space version \cite{Neiman:2015wma} of the Vasiliev equations, where the decomposition into right-handed/left-handed spinor spaces depends on the spacetime coordinate $x^\mu$ via the projectors $P(\pm x)$, as in section \ref{sec:geometry:spinors}. 

In the Vasiliev equations, the HS algebra \eqref{eq:star} is extended to include an additional twistor $Z^a$, with star-products:
\begin{align}
	Y^a\star Z^b = Z^b\star Y^a = Y^aZ^b - iI^{ab} \ ; \quad Z^a\star Z^b = Z^aZ^b - iI^{ab} \ .
\end{align}
The right-handed and left-handed spinor delta-functions w.r.t. $Y$ and $Z$ combine into the internal Klein operators:
\begin{align}
	\delta_{\pm x}(Y)\star\delta_{\pm x}(Z) = e^{iYP_{\pm x}Z} \equiv \kappa_{\pm x} \ ,
\end{align}
which satisfy:
\begin{align}
	\kappa_{\pm x}\star\kappa_{\pm x} = 1 \ ; \quad \kappa_{\pm x}\star\kappa_{\mp x} = e^{iYZ} \equiv \calK \ .
\end{align}
In the most economical formulation of Vasiliev's equations, the fundamental master fields are the connection 1-form $W(x;Y,Z) \equiv W_\mu(x;Y,Z)dx^\mu$, whose vacuum value in the embedding-space formulation is $W=0$, and the twistor $S_a(x;Y,Z)$, whose vacuum value is $S_a = Z_a$. The curvature master field $C(x;Y)$ from section \ref{sec:geometry:Penrose} becomes the linearized approximation to a non-linear field $B(x;Y,Z)$, which can be defined in terms of the ``potential'' $S_a$ as:
\begin{align}
	B \equiv e^{i\phi}\left(\frac{i}{2}P^{ab}(x)S_a\star S_b - 1\right)\star\kappa_x \ . \label{eq:B}
\end{align}
Here, the phase $\phi$ captures the 1-parameter freedom in the interactions of HS Gravity. In this subsection, we allow a general $\phi$. In the rest of this paper, we consider the type-A and type-B theories, which correspond to $\phi=0$ (the distinction between type-A and type-B will be captured separately). We take the fields to be integer-spin, which means that $W,B$ are even functions of $(Y,Z)$, while $S_a$ is an odd function. The field equations read \cite{Neiman:2015wma}:
\begin{gather}
	dW + W\star W = \frac{i}{16}\,dx^a{}_c\,dx^{cb}\,S_a\star S_b \ ; \label{eq:dW} \\
	dS_a + W\star S_a - S_a\star W = -\frac{1}{2}x^b{}_c\,dx^c{}_a\,S_b \ ; \\
	\left(\frac{i}{2}P^{ab}(-x)S_a\star S_b - 1\right) = \eta\,e^{2i\phi}\left(\frac{i}{2}P^{ab}(x)S_a\star S_b - 1\right)\star\calK \ . \label{eq:S_S}
\end{gather}
Here, $\eta$ is a sign factor, which for the moment is redundant with the phase $\phi$. For $\phi=0$, the value $\eta = \pm 1$ corresponds to type-A/type-B theory, respectively. Alternatively, we can upgrade $\eta$ into an \emph{operator with eigenvalues} $\pm 1$, which commutes with $(Y,Z)$, and enters as yet another argument in the master fields $W,S_a,B$. In that case, the equations \eqref{eq:dW}-\eqref{eq:S_S} describe two non-interacting bosonic subsectors (type-A and type-B for $\phi=0$), one for each eigenspace of $\eta$. In AdS, these would comprise the bosonic sector of the $\calN=2$ supersymmetric theory. 

Let's now construct reality conditions based on the ``new'' complex conjugation \eqref{eq:new_conjugation}. First, we extend \eqref{eq:new_conjugation} in the simplest way to functions of both $Y^a$ and $Z^a$:
\begin{align}
	f^*(Y,Z) \equiv \overline{f(-i\bar Y,-i\bar Z)} \ . \label{eq:new_conjugation_Z}
\end{align}
The virtue of this definition is that it continues to preserve the star product, just as in \eqref{eq:star_order} (again, without reversing its order). We also have:
\begin{align}
	\kappa^*_{\pm x} = \kappa_{\mp x} \ ; \quad \calK^* = \calK \ , \label{eq:kappa_reality}
\end{align}
where we assumed that the spacetime coordinates $x^\mu$ are real. We now define the following reality conditions on the master fields $W$ and $S_a$ (again, at real $x$):
\begin{align}
	W^\star = W \ ; \quad S_a^\star = -iS_a \ . \label{eq:W_S_reality}
\end{align}
It's easy to verify that the reality conditions \eqref{eq:W_S_reality} are consistent with the Vasiliev equations \eqref{eq:dW}-\eqref{eq:S_S}. We simply need to check that, assuming the reality condition \eqref{eq:W_S_reality}, eqs. \eqref{eq:dW}-\eqref{eq:S_S} are invariant under the complex conjugation \eqref{eq:new_conjugation_Z}. Using the identities \eqref{eq:star_order},\eqref{eq:kappa_reality}, this check is straightforward.

Finally, for the curvature master field \eqref{eq:B}, the reality conditions \eqref{eq:W_S_reality} together with the field equation \eqref{eq:S_S} imply:
\begin{align}
	B^* = \eta B \ , \label{eq:B_reality}
\end{align}
which we immediately recognize as the non-linear extension of \eqref{eq:reality_star}. We conclude that the linearized reality condition \eqref{eq:reality_star} on the curvatures indeed extends to the full Vasiliev equations, via the reality conditions \eqref{eq:W_S_reality} on the potentials.

\subsection{``New'' reality condition vs. fermions} \label{sec:reality:fermions}

Let us now show that the ``new'' reality condition \eqref{eq:reality_star} is inconsistent with fermions, and thus with supersymmetry. In its simplest form, this observation is immediate. Indeed, applying \eqref{eq:reality_star} twice and recalling \eqref{eq:star_identities}, we get $C(x;-Y) = C(x;Y)$, i.e. a restriction to integer spins. This is in contrast with the standard reality condition \eqref{eq:C_dagger}, which, when applied twice, gives the identity $C(x;-Y) = \delta(Y)\star C(x;Y)\star\delta(Y)$.

Now, this consideration may be considered naive. This is because, at the level of the Vasiliev equations, one doesn't introduce fermions merely by allowing odd functions of $Y$. Instead, one extends the star-product algebra by either external Klein operators \cite{Vasiliev:1990en,Vasiliev:1992av}, or Grassmann coordinates \cite{Engquist:2002vr,Sezgin:2012ag}. However, as we'll now see, these do not change the final result. 

First, let's try to construct the supersymmetric theory using external Klein operators, as in \cite{Vasiliev:1990en,Vasiliev:1992av}. In this language, the master fields become unconstrained functions of $(Y,Z)$, while the sign factor $\eta$ in \eqref{eq:S_S},\eqref{eq:B_reality} is upgraded into a Klein operator $K$ that anti-commutes with $(Y,Z)$. To maintain consistency, $K$ must inherit from $\eta$ the properties $K^2 = 1$ and $K^* = K$. Also, we must extend the complex conjugation \eqref{eq:new_conjugation_Z} to functions of $K$ so that it still preserves the star-product as in \eqref{eq:star_order}:
\begin{align}
	\big(f(Y,Z)K\big)^* = f^*(Y,Z)K \ .
\end{align}
But now we end up in the same place as before: applying the reality condition \eqref{eq:B_reality} twice, we get $B(-Y,-Z) = B(Y,Z)$ (suppressing irrelevant arguments), i.e. again a restriction to integer spins. This is again in contrast to the standard reality condition, which is based on a complex conjugation $f\to f^\dagger$ that reverses the order of star-products, such that applying it twice yields the identity $B(-Y,-Z) = KB(Y,Z)K$.

Finally, let's try the approach of Grassmann coordinates \cite{Engquist:2002vr,Sezgin:2012ag}. We introduce two coordinates $\theta_p = (\theta_1,\theta_2)$, which commute with $(Y,Z)$, and satisfy:
\begin{align}
	\{\theta_p,\theta_q\} = 2\delta_{pq} \ . \label{eq:theta_anticommute}
\end{align}
The master fields $W,B$ are now even functions of the entire set $(Y,Z,\theta)$, with $S_a$ an odd function of the same. The sign factor $\eta$ in \eqref{eq:S_S},\eqref{eq:B_reality} is now upgraded into an operator $\Gamma$, constructed from the $\theta$'s as:
\begin{align}
	\Gamma = \frac{i}{2}\epsilon_{pq}\theta_p\theta_q \ , \label{eq:Gamma}
\end{align}
where the prefactor is chosen so as to inherit from $\eta$ the property $\Gamma^2 = 1$. Now, let's try to define the complex conjugation of the $\theta$'s as:
\begin{align}
	\theta^*_p = A_{pq}\theta_q \ , \label{eq:theta}
\end{align}
for some matrix $A_{pq}$. Again, for functions of both $(Y,Z)$ and $\theta$, we must define the complex conjugation so that the star-product remains preserved as in \eqref{eq:star_order}. We then have the following requirements from the matrix $A_{pq}$:
\begin{enumerate}
	\item The complex conjugation should preserve the anti-commutation relations \eqref{eq:theta_anticommute}, which requires $AA^T = 1$.
	\item $\Gamma$ from \eqref{eq:Gamma} should be real, which requires $\det A = -1$.
	\item For fermions to be consistent with the reality condition, the complex conjugation of the $\theta$'s should square to $-1$, to balance the same property of the twistors $(Y,Z)$. This requires $AA^* = -1$.
\end{enumerate}
Our observation is now simply that no such matrix $A$ exists. Again, this is in contrast with the standard reality conditions based on $f\to f^\dagger$. There, due to the order of products getting reversed under complex conjugation, the reality of $\Gamma$ requires $\det A = 1$ rather than $\det A = -1$. One then has the solution $A_{pq} = \epsilon_{pq}$ from \cite{Sezgin:2012ag}. 

In summary, the ``new'' reality condition \eqref{eq:reality_star}, which will prove to be the right one for odd spins in dS/CFT, is consistent with the bosonic Vasiliev equations, but not with fermions or supersymmetry.

\section{Hermitian norm for HS (super)multiplet in de Sitter} \label{sec:Hermitian}

In this section, we present for the first time a simple HS-algebraic construction of the Hilbert space for a single-particle HS (super)multiplet in de Sitter. In section \ref{sec:Hermitian:basis}, we describe the states in this Hilbert space and the operators that act on them. In section \ref{sec:Hermitian:norm}, we describe the Hermitian norm on states, and the Hermitian conjugation of operators, using the ``new'' complex conjugation from \cite{Chang:2013afa} and section \ref{sec:reality}. In section \ref{sec:Hermitian:symmetry}, we describe the action of HS (super)symmetry on the states.

\subsection{Wavefunction basis} \label{sec:Hermitian:basis}

The states in the HS single-particle Hilbert space are described by \emph{positive-frequency} solutions \eqref{eq:Penrose} to the free massless field equations:
\begin{align}
	C_\Psi(x;Y) = iF_\Psi(Y)\star\delta_x(Y) \ . \label{eq:Penrose_Psi}
\end{align}
The restriction to positive frequency can be expressed in spacetime as $C_\Psi(x;Y)$ being regular when analytically continued to $EAdS_4^+$. Alternatively, it can be expressed in twistor space as $F_\Psi(Y)$ being regular on $\bbT^+$. Since these are not algebraic conditions, it will be useful to introduce an explicit basis of appropriate functions. For that purpose, let us choose an arbitrary origin point $x_E\in EAdS_4^+$:
\begin{align}
	x_E^\mu = (-i,0,0,0,0) \ . \label{eq:x_E}
\end{align}
Our wavefunction basis is now given by the Taylor expansion of $C_\Psi(x_E;Y)$ w.r.t. the right-handed and left-handed parts of its twistor argument:
\begin{align}
	C_\Psi(x_E;Y) = \sum_{m,n=0}^\infty \frac{1}{m!n!}\,\Psi^{(m,n)}_{a_1\dots a_m b_1\dots b_n}\,y_{(x_E)}^{a_1}\dots y_{(x_E)}^{a_m}\,y_{(-x_E)}^{b_1}\dots y_{(-x_E)}^{b_n} \ . \label{eq:C_Psi}
\end{align}
Via \eqref{eq:packaging},\eqref{eq:unfolding}, the spin-tensor coefficients $\Psi^{(m,n)}$ simply describe the values and independent derivatives of the curvatures of all spins at $x_E$. Each spin $s$ appears once, with $\Psi^{(2s+k,k)}$ and $\Psi^{(k,2s+k)}$ encoding the $k$'th derivative of its right-handed and left-handed curvature, respectively. A priori, half-integer spins are included, making an HS \emph{super}multiplet, whose supersymmetry we will describe in the next subsection. Since we're discussing single-particle states, the distinction between Bose and Fermi statistics does not appear.

When we evolve $C_\Psi(x;Y)$ away from $x=x_E$, each term in \eqref{eq:C_Psi} becomes a positive-frequency solution of the field equations, with $\Psi^{(m,n)}$ describing its behavior under $SO(4)$ rotations around $x_E$. The twistor function $F_\Psi(Y)$ corresponding to \eqref{eq:C_Psi} is given by:
\begin{align}
	F_\Psi(Y) = -i\sum_{m,n=0}^\infty \frac{i^m}{m!n!}\,\Psi_{(m,n)}^{a_1\dots a_m}{}_{b_1\dots b_n}\,y_{(-x_E)}^{b_1}\dots y_{(-x_E)}^{b_n}\frac{\del^m}{\del y^{a_1}_{(x_E)}\dots\del y^{a_m}_{(x_E)}}\,\delta_{x_E}(Y) \ . \label{eq:F_Psi}
\end{align}
We can see that each term in \eqref{eq:F_Psi} is positive-frequency, since it is singular in twistor space only on the support of $\delta_{x_E}(Y)$, i.e. only on the subspace $P(-x_E)$, which lies entirely in $\bbT^-$.

Let's now discuss linear \emph{operators} $\ket{\Psi}\to\hat O\ket{\Psi}$ acting on the Hilbert space of states \eqref{eq:C_Psi}-\eqref{eq:F_Psi}. Since the twistor function $F_\Psi(Y)$ is a more covariant object than the master field $C_\Psi(x_E;Y)$ or its Taylor coefficients $\Psi^{(m,n)}$, it will be most convenient to express the action of operators on $F_\Psi(Y)$. The most general operator can be composed from multiplication by $Y^a$ and differentiation $\del/\del Y^a$. In HS algebra, these can be expressed respectively as the anticommutator $\frac{1}{2}\{Y^a,F_\Psi(Y)\}_\star$ and commutator $\frac{i}{2}[Y_a,F_\Psi(Y)]_\star$. Thus, \emph{every operator can be expressed in terms of the HS star-product}. Of particular interest are the operators given by a star-product with an arbitrary (polynomial) function $G(Y)$ on the left or on the right:
\begin{align}
	\widehat{G(Y)}_L \ &: \quad F_\Psi(Y) \to G(Y)\star F_\Psi(Y) \ ; \label{eq:O_L} \\
	\widehat{G(Y)}_R \ &: \quad F_\Psi(Y) \to F_\Psi(Y)\star G(Y) \ . \label{eq:O_R}
\end{align}

\subsection{Hermitian norm} \label{sec:Hermitian:norm}

Having discussed the linear structure of the Hilbert space, we turn to define its Hermitian norm. We can derive this in a couple of steps from the real norm $\bar UU = \bar U_a U^a$ on twistor space; note that this structure is specific to de Sitter, with no analogue in Lorentzian AdS. An immediate obstacle is that the norm $\bar UU$ is only \emph{pseudo-}Hermitian, having mixed signature $(+,+,-,-)$. However, this is easy to correct once we fix a Euclidean origin point $x^\mu_E \in EAdS_4^+$, as in \eqref{eq:x_E}. We then see that $\bar U U$ decomposes into a positive-definite norm $\bar UP(x_E)U$ on the right-handed spinors $u_{x_E}\in P(x_E)\subset \bbT^+$, plus a \emph{negative-}definite norm $\bar UP(-x_E)U$ on the left-handed spinors $u_{-x_E}\in P(-x_E)\subset \bbT^-$. We can thus construct a positive-definite norm, namely the standard norm on $SO(4)$ spinors, by flipping the sign of the left-handed part, as:
\begin{align}
	\bar UP(x_E)U - \bar UP(-x_E)U = i\bar Ux_E U \ .
\end{align}
We can apply this norm to the wavefunction coefficients \eqref{eq:C_Psi} to define a positive-definite Hermitian inner product as:
\begin{align}
	\braket{\Psi|\Psi} = \sum_{m,n=0}^\infty \frac{(-1)^n}{m!n!}\,\bar\Psi^{(m,n)}_{a_1\dots a_mb_1\dots b_n}\,\Psi_{(m,n)}^{a_1\dots a_mb_1\dots b_n} \ . \label{eq:Psi^2_raw}
\end{align}
Note that the inner product \eqref{eq:Psi^2_raw} decomposes into orthogonal subspaces for the different helicities $h=\frac{1}{2}(m-n)$, as required by $SO(1,4)$ symmetry.

Now, recall that in HS algebra, flipping the sign of the left-handed or right-handed spinor space at a bulk point can be accomplished via the star-product conjugation \eqref{eq:bulk_delta_flip}. Using \eqref{eq:star_diff},\eqref{eq:str}, we can now package the Hermitian norm \eqref{eq:Psi^2_raw} into an HS-algebraic traced product:
\begin{align}
	\braket{\Psi|\Psi} = \tr_\star\big(C_\Psi^*(x_E;Y)\star\delta_{x_E}(Y)\star C_\Psi(x_E;Y)\star\delta_{x_E}(Y)\big) \ , \label{eq:Psi^2_C}
\end{align}
where, crucially, we used the ``new'' complex conjugation \eqref{eq:new_conjugation} to absorb the factors of $i$ in the star-product formula \eqref{eq:star_diff}. The Hermitian norm \eqref{eq:Psi^2_C} is positive-definite for all integer and half-integer spins. If we were to replace the complex conjugation $C_\Psi^*$ with the standard one $C_\Psi^\dagger$, the result would become negative-definite for odd spins, and imaginary for half-integer ones (while remaining unchanged for even spins). 

Finally, we can use the Penrose transform \eqref{eq:Penrose_Psi} to express \eqref{eq:Psi^2_C} in terms of the twistor function $F_\Psi(Y)$:
\begin{align}
	\braket{\Psi|\Psi} = -\tr_\star\big(F_\Psi^*(Y)\star F_\Psi(Y)\big) \ . \label{eq:Psi^2_F}
\end{align}
It is now manifest that the Hermitian norm $\braket{\Psi|\Psi}$ is HS-symmetric, and doesn't depend on our choice of Euclidean origin point $x_E$. In particular, this implies that the norm is symmetric under the full de Sitter group $SO(1,4)$, rather than just the $SO(4)$ rotations around $x_E$ that are manifest in \eqref{eq:Psi^2_raw}. Note that for fermions with spin $s\geq \frac{3}{2}$, the precise correspondence between spacetime fields (in our context, $C_\Psi(x;Y)$) and unitary representations of the de Sitter group (in our context, the Hermitian norm $\braket{\Psi|\Psi}$), is by itself a recent development \cite{Letsios:2023qzq,Letsios:2022tsq,Letsios:2023awz}.

As always for free fields (which inherit this property from harmonic oscillators), the single-particle Hermitian inner product $\braket{\Phi|\Psi}$ can be recast as the symplectic form of the free field theory, evaluated on the creation and annihilation operators of particles in the states $\ket{\Phi},\ket{\Psi}$. In terms of spacetime fields of individual spins, this symplectic form is nothing but the Klein-Gordon inner product. When evaluated on the spacetime's boundary, it becomes the boundary 2-point function (see section \ref{sec:boundary:2_point} below). For individual spins, these closely-related structures are all $SO(1,4)$-invariant and unique. Eq. \eqref{eq:Psi^2_F} then combines the individual spins in the essentially unique HS-invariant way. Indeed, it's the only available quadratic HS-algebraic expression, up to the discrete choice of complex conjugation, which, as we saw, is driven by the positivity requirement. See e.g. \cite{David:2020fea} for a discussion that goes from the symplectic form, via the 2-point function, to a quadratic star-product expression similar to \eqref{eq:Psi^2_F}. Again, what's new in our present treatment is the specific choice of complex-conjugation structure in \eqref{eq:Psi^2_F}, which ensures a positive-definite norm for all spins in de Sitter.

The norm \eqref{eq:Psi^2_F} defines a Hermitian conjugation $\hat O\to\hat O^\dagger$ for operators acting on the Hilbert space, via $\braket{\hat O^\dagger\Phi|\Psi} = \braket{\Phi|\hat O\Psi}$. For instance, the Hermitian conjugates of the special operators \eqref{eq:O_L}-\eqref{eq:O_R} read:
\begin{align}
	\left(\widehat{G(Y)}_L\right)^\dagger \ &: \quad F_\Psi(Y) \to F_\Psi(Y)\star G^*(-Y) \ ; \label{eq:O_L_dagger} \\
	\left(\widehat{G(Y)}_R\right)^\dagger \ &: \quad F_\Psi(Y) \to G^*(Y)\star F_\Psi(Y) \ . \label{eq:O_R_dagger}
\end{align}
Another example is the fermion-parity operator $F_\Psi(Y) \to F_\Psi(-Y)$, which is Hermitian, as can be seen e.g. by combining \eqref{eq:O_L_dagger}-\eqref{eq:O_R_dagger} and setting $G(Y) = \delta(Y)$.

\subsection{HS (super)symmetry} \label{sec:Hermitian:symmetry}

Let's now explicitly describe the HS symmetry of the above single-particle Hilbert space. Here, we'll consider the case of the \emph{super}multiplet, namely with no restriction of \eqref{eq:C_Psi}-\eqref{eq:F_Psi} to integer spins. Accordingly, the HS (super)symmetry generators are polynomial functions $\varepsilon(Y)$ of arbitrary order. The bosonic sector can always be recovered by restricting the wavefunctions \eqref{eq:C_Psi}-\eqref{eq:F_Psi} and symmetry generators $\varepsilon(Y)$ to even functions of $Y$. Quadratic functions $\varepsilon(Y) = M_{ab}Y^a Y^b$ generate the $SO(1,4)$ de Sitter isometry group. Linear functions $\varepsilon(Y) = M_a Y^a$ are supersymmetry (SUSY) generators, related to the $SO(1,4)$ generators via the basic anticommutator (note that for anticommutators, we can take both factors equal without loss of generality):
\begin{align}
 \left\{M_a Y^a,M_b\,Y^b\right\}_\star = (2M_a M_b) Y^a Y^b \ . \label{eq:basic_anti}
\end{align}
Higher polynomials $\varepsilon(Y)$ encode higher-spin (super)symmetries.  To describe both bosonic and fermionic states and symmetries, we define a graded star-product commutator:
\begin{align}
	\big[f(Y),g(Y)\big\}_\star \equiv f(Y)\star g(Y) \mp g(Y)\star f(Y) \ ,
\end{align}
where the sign is ``+'' iff both $f(Y)$ and $g(Y)$ are odd functions. The HS symmetry on the Hilbert space is then defined by the operators:
\begin{align}
	\widehat{\varepsilon(Y)} \ : \quad F_\Psi(Y) \to \Big[\varepsilon(Y),F_\Psi(Y)\Big\}_\star \ , \label{eq:generator}
\end{align}
whose graded algebra reproduces that of the star-product, i.e. if $\big[\varepsilon_1(Y),\varepsilon_2(Y)\big\}_\star = \varepsilon_3(Y)$, then also $\big[\widehat{\varepsilon_1(Y)},\widehat{\varepsilon_2(Y)}\big\} = \widehat{\varepsilon_3(Y)}$. 

The action of the fermionic symmetries in \eqref{eq:generator} is \emph{chiral} -- a possibly surprising feature that is inherited from the chirality of the Penrose transform \eqref{eq:Penrose_Psi}. An easy way to see the generators' chirality is to consider their effect on the helicity of our particle state. Recall from \eqref{eq:homogeneity} that the helicity is linearly related to the degree of homogeneity of the twistor function $F_\Psi(Y)$. For a bosonic generator $\varepsilon(Y)$ of even degree $2n$, the star-commutator $[\varepsilon,F_\Psi]_\star$ shifts the helicity $h$ by (a superposition of) the integers $(1-n,\dots,n-1)$. This range of shifts is symmetric around 0, indicating that the bosonic generators aren't chiral. Now, for a fermionic generator  $\varepsilon(Y)$ of odd degree $2n+1$, the star-commutator $[\varepsilon,F_\Psi]_\star$ shifts a bosonic helicity by the half-integers $(-n+\frac{1}{2},n+\frac{1}{2})$, while the anti-commutator $\{\varepsilon,F_\Psi\}_\star$ shifts a fermionic helicity by the half-integers $(-n-\frac{1}{2},n-\frac{1}{2})$. Since these ranges aren't symmetric around 0, we conclude that the action of fermionic generators (on both bosonic and fermionic states) is chiral.

The master field $C_\Psi(x_E;Y)$ transforms under the HS generators \eqref{eq:generator} as:
\begin{align}
	C_\Psi(x_E;Y) \ \to \ \varepsilon(Y)\star C_\Psi(x_E;Y) \mp C_\Psi(x_E;Y)\star\delta_{x_E}(Y)\star\varepsilon(Y)\star\delta_{x_E}(Y) \ . \label{eq:generator_C}
\end{align}
Thus, we say that $F_\Psi(Y)$ transforms in the adjoint representation, while $C_\Psi(x_E;Y)$ transforms in the ``twisted adjoint''. Note that, via \eqref{eq:star_diff},\eqref{eq:bulk_delta_flip}, the transformation \eqref{eq:generator_C} sends polynomials to polynomials. This means that HS symmetry generators map any positive-frequency basis element from \eqref{eq:C_Psi} into a finite superposition of other such elements.

The Hermitian conjugates of the generators \eqref{eq:generator} under the Hilbert-space norm \eqref{eq:Psi^2_F} can be read off from \eqref{eq:O_L_dagger}-\eqref{eq:O_R_dagger} as:
\begin{align}
	\left(\widehat{\varepsilon(Y)}\right)^\dagger \ : \quad F_\Psi(Y) \to \Big[F_\Psi(Y), \varepsilon^*(-Y)\Big\}_\star \ . \label{eq:generator_dagger}
\end{align}
For the bosonic generators, where the graded star-commutator $[,\}_\star$ is just a commutator $[,]_\star$, this implies the mundane result $\big(\widehat{\varepsilon(Y)}\big)^\dagger = -\widehat{\varepsilon^*(Y)}$. For the fermionic generators, including the linear SUSY generators $\varepsilon(Y) = M_a Y^a$, the result is more exotic: the Hermitian conjugate \eqref{eq:generator_dagger} does not belong to the original class of operators \eqref{eq:generator}. Note that the Hermitian-conjugate fermionic generators are again chiral, \emph{with the same chirality} as the original fermionic generators. In particular, their effect on helicities is the same as for the original \eqref{eq:generator}. 

If we try to anti-commute the original fermionic generators $\varepsilon(Y)$ with their Hermitian conjugates $\big(\widehat{\varepsilon(Y)}\big)^\dagger$, we get a rather complicated operator that isn't one of the original bosonic generators \eqref{eq:generator}: 
\begin{align}
  	  &\Big\{\widehat{\varepsilon(Y)},\left(\widehat{\varepsilon(Y)}\right)^\dagger\Big\} \ : \label{eq:mixed_fermion} \\ 
	     &F_\Psi(Y) \ \to \ 2\big(\varepsilon(Y)^*\star F_\Psi(Y)\star\varepsilon(Y) - \varepsilon(Y)\star F_\Psi(Y)\star\varepsilon^*(Y)\big) + \Big\{\big[\varepsilon(Y),\varepsilon^*(Y)]_\star,F_\Psi(-Y)\Big\}_\star \ . \nonumber
\end{align}
The situation is thus as follows. Firstly, we have the bosonic HS symmetry algebra with generators $\widehat{\varepsilon(Y)}$ parameterized by even functions $\varepsilon(Y)$, which can be chosen anti-Hermitian by setting $\varepsilon(Y) = \varepsilon^*(Y)$. Second, this can be expanded into a \emph{chiral} superalgebra by allowing odd functions $\varepsilon(Y)$, with the peculiar property that the Hermitian conjugates $\big(\widehat{\varepsilon(Y)}\big)^\dagger$ form a \emph{different copy} of the superalgebra, which intersects with the original only on the bosonic sector. If we try to combine the superalgebra with its Hermitian conjugate, we end up with new bosonic generators that weren't part of the original HS symmetry. 

Zooming out, the takeaway of this section is that we can define an HS single-particle Hilbert space in de Sitter, with positive-definite Hermitian norm for all spins \eqref{eq:Psi^2_raw}-\eqref{eq:Psi^2_F} constructed out of the HS star-product and the complex conjugation \eqref{eq:new_conjugation}. This Hilbert space forms a representation of the bosonic HS symmetry algebra, and also of its superalgebra extension, up to the above peculiarities. The fermionic generators and states are probably not relevant in the bigger picture, since in the end supersymmetric HS Gravity in de Sitter doesn't appear to exist. We already saw one piece of evidence for this: when the complex conjugation \eqref{eq:new_conjugation} that yields a positive-definite norm \eqref{eq:Psi^2_raw}-\eqref{eq:Psi^2_F} for HS \emph{particles} is used to define a reality condition \eqref{eq:reality_star} for HS \emph{fields}, that reality condition only supports integer spins.

\subsection{The SUSY sector: linear and quadratic generators}

In this subsection, we consider more closely the ``SUSY sector'' of symmetries, namely the linear fermionic generators $\varepsilon(Y) = M_a Y^a$, and the quadratic bosonic generators on which their algebra closes. It will turn out that our construction nicely generalizes the one found in \cite{Higuchi:2025pbc} for a chiral SUGRA supermultiplet in de Sitter.

The SUSY generators $\varepsilon(Y) = M_a Y^a$ act on bosonic (fermionic) states respectively as a star-product commutator (anti-commutator) with $Y^a$; this decreases (increases) the degree of homogeneity of $F_\Psi(Y)$ by 1, which raises (lowers) the helicity by $1/2$. Thus, under these SUSY generators, the different helicities decompose into chiral doublets $(h,h+\frac{1}{2})$, where $h$ is an integer. The effect on helicities of the Hermitian-conjugate operators \eqref{eq:generator_dagger} is the same. This is precisely the situation found in \cite{Higuchi:2025pbc}, for the particular helicity doublet $(-2,-\frac{3}{2})$. 

The anti-commutators of the original SUSY operators \eqref{eq:generator} and their Hermitian conjugates \eqref{eq:generator_dagger} with themselves read:
\begin{align}
  \Big\{\widehat{\varepsilon(Y)},\widehat{\varepsilon(Y)}\Big\} \ &: \quad F_\Psi(Y) \to 8iM_a M^b\,Y^a\frac{\del F_\Psi(Y)}{\del Y^b} \ ; \label{eq:Q_Q} \\
  \Big\{\left(\widehat{\varepsilon(Y)}\right)^\dagger,\left(\widehat{\varepsilon(Y)}\right)^\dagger\Big\} \ &: \quad F_\Psi(Y) \to 8i\bar M_a\bar M^b\, Y^a\frac{\del F_\Psi(Y)}{\del Y^b} \ . \label{eq:Qbar_Qbar}
\end{align}
The real combination of \eqref{eq:Q_Q} and \eqref{eq:Qbar_Qbar} generates the $SO(1,4)$ de Sitter algebra, parameterized by real symmetric matrices $M^{\text{symm}}_{ab} = M_aM_b + \bar M_a\bar M_b$. These correspond of course to the quadratic generators $\sim M^{\text{symm}}_{ab}Y^a Y^b$ in \eqref{eq:generator}. 

Now, consider the anti-commutator \eqref{eq:mixed_fermion} between $\widehat{\varepsilon(Y)}$ and its Hermitian conjugate:
\begin{align}
	\begin{split}
	&\Big\{\widehat{\varepsilon(Y)},\left(\widehat{\varepsilon(Y)}\right)^\dagger\Big\} \ : \\
	&\qquad F_\Psi(Y) \ \to \ 4\bar M_aM^a\big(F_\Psi(-Y) - F_\Psi(Y)\big) + 4\big(M_a\bar M^b - \bar M_a M^b \big)Y^a\frac{\del F_\Psi(Y)}{\del Y^b} \ . 
	\end{split} \label{eq:QQbar_raw}
\end{align}
Let us denote the \emph{traceless} part of the real antisymmetric matrix $2M_{[a}\bar M_{b]}$ as:
\begin{align}
	M^{\text{skew}}_{ab} \equiv 2M_{[a}\bar M_{b]} - \frac{1}{2}\bar M_c M^c I_{ab} \ .
\end{align}
Then, using \eqref{eq:homogeneity}, the action of \eqref{eq:QQbar_raw} on a (boson, fermion) doublet with helicities $(h,h+\frac{1}{2})$ reads:
\begin{align}
	\begin{split}
	&\Big\{\widehat{\varepsilon(Y)},\left(\widehat{\varepsilon(Y)}\right)^\dagger\Big\} \ : \\
	&\qquad F_\Psi(Y) \ \to \ -4(M^{\text{skew}})^a{}_b\,Y^b\frac{\del F_\Psi(Y)}{\del Y^a} + 2\bar M_a M^a F_\Psi(Y)\times\left\{
	\begin{array}{ll}
		2h+2 & \quad \text{boson} \\
		2h-1 & \quad \text{fermion}
	\end{array} \right. \ . 
	\end{split} \label{eq:QQbar}
\end{align}
Here, the first term has the same form as the $SO(1,4)$ generators \eqref{eq:Q_Q}-\eqref{eq:Qbar_Qbar}, but with an imaginary anti-symmetric traceless matrix $iM^{\text{skew}}_{ab}$ in place of the real symmetric one $M^{\text{symm}}_{ab}$. Together, the two span the algebra of all \emph{Hermitian traceless} matrices $M_{ab}$, which is nothing but the conformal algebra $SU(2,2)\cong SO(2,4)$ (recalling that ``Hermitian'' here refers to the pseudo-Hermitian norm $\bar U_a U^a$ of section \ref{sec:geometry:spinors}). 

The second term in \eqref{eq:QQbar} is a $U(1)$ generator. Again, this is the same situation observed for the $(-2,-\frac{3}{2})$ doublet in \cite{Higuchi:2025pbc}, where a bosonic algebra $SO(2,4)\times U(1)$ was identified. As a nice cross-check, the ratio of the $U(1)$ charges for the boson and fermion in \eqref{eq:QQbar} in the case $h=2$ is $2/5$, precisely as observed in \cite{Higuchi:2025pbc}.

\section{New reality conditions for the boundary vector model} \label{sec:boundary}

In this section, we construct CFT duals to HS Gravity in de Sitter, extending the minimal type-A construction of \cite{Anninos:2011ui} to all integer spins and to type-B. The non-trivial task is to choose the right reality conditions when adapting the well-known AdS/CFT duals to de Sitter, so as to get 2-point functions with the correct signs. Motivated by the result of section \ref{sec:Hermitian:norm}, where the complex conjugation \eqref{eq:new_conjugation} gives positive norms to all spins, we expect that the correct reality conditions on bulk fields and twistor functions are those in \eqref{eq:reality_star}. Recalling that these are related to the standard reality conditions \eqref{eq:C_dagger}-\eqref{eq:F_dagger} by a real$\leftrightarrow$imaginary flip for odd spins (c.f. a similar observation in \cite{Hertog:2017ymy}), we set out to engineer such a flip in the boundary theory. In section \ref{sec:boundary:2_point}, we will verify our ``guess'' by computing the 2-point functions in HS algebra.

\subsection{Type-A Euclidean AdS/CFT} \label{sec:boundary:type_A_EAdS}

We start by describing the known boundary dual of type-A HS Gravity in Euclidean AdS. This is given by a free $U(N)$ vector model, with action:
\begin{align}
	S_{\text{CFT}} = \int d^3\mathbf{r}\,(\boldsymbol{\nabla}\bar\phi_I)\cdot(\boldsymbol{\nabla}\phi^I) \ . \label{eq:EAdS_type_A}
\end{align}
Here, for simplicity, we use the flat conformal frame \eqref{eq:Poincare} for the 3d boundary, with $\boldsymbol{\nabla}\equiv \del/\del\mathbf{r}$. The $\phi^I$ are scalar fields with complex conjugates $\bar\phi_I$, where $I=1,\dots,N$ is a color index. The single-trace primaries of the vector model \eqref{eq:EAdS_type_A} are a tower of currents with all integer spins $s$ (including an ``honorary current'' of spin 0). These currents are traceless symmetric tensors $J^{(s)}_{i_1\dots i_s}(\mathbf{r})$, which can be packaged conveniently as:
\begin{align}
	J^{(s)}(\mathbf{r};\boldsymbol{\lambda}) = \lambda^{i_1}\dots\lambda^{i_s}J^{(s)}_{i_1\dots i_s}(\mathbf{r}) \ , \label{eq:J_packaging}
\end{align}
 with $\boldsymbol{\lambda}$ a complex null 3d vector $\boldsymbol{\lambda}^2=0$. All of the local spin-$s$ currents $J^{(s)}(\mathbf{r};\boldsymbol{\lambda})$ can be constructed from the bilocal single-trace operator $\calO(\mathbf{r},\mathbf{r'}) \equiv \bar\phi_I(\mathbf{r'})\phi^I(\mathbf{r})$, as:
\begin{align}
	J^{(s)}(\mathbf{r};\boldsymbol{\lambda}) = i^s\sum_{k=0}^s (-1)^k \binom{2s}{2k} (\boldsymbol{\lambda\cdot\nabla})^k (\boldsymbol{\lambda\cdot\nabla'})^{s-k} \, \bar\phi_I(\mathbf{r'})\phi^I(\mathbf{r}) \Big|_{\mathbf{r'=r}} \ , \label{eq:J_EAdS_type_A}
\end{align}
where $\boldsymbol{\nabla'}\equiv \del/\del\mathbf{r'}$. The $n$-point correlators $\langle J^{(s_1)}(\mathbf{r_1};\boldsymbol{\lambda_1})\dots J^{(s_n)}(\mathbf{r_n};\boldsymbol{\lambda_n}) \rangle$ can all be obtained by applying the differential operators \eqref{eq:J_EAdS_type_A} to the fundamental correlator of bilocals:
\begin{align}
	\big\langle \calO(\mathbf{r_1},\mathbf{r'_1})\dots \calO(\mathbf{r_n},\mathbf{r'_n}) \big\rangle = \frac{N}{(4\pi)^n} \left(\prod_{p=1}^n \frac{1}{|\mathbf{r_{p+1} - r'_p}|} + \text{permutations}\right) \ . \label{eq:correlator_EAdS_type_A}
\end{align}
Here, the product is cyclic, i.e. $\mathbf{r_{n+1}}\equiv\mathbf{r_1}$, and ``+ permutations'' denotes a sum over the $(n-1)!$ cyclically inequivalent permutations of $(1,\dots,n)$. Each permutation corresponds to a 1-loop Feynman diagram in the boundary theory. The factors of $1/(4\pi|\mathbf{r_{p+1}-r'_p}|)$ in \eqref{eq:correlator_EAdS_type_A} are simply the propagators inverse to the kinetic operator $\boldsymbol{\nabla\cdot\nabla}$ in \eqref{eq:EAdS_type_A}. Here and throughout, we take $\langle\dots\rangle$ to denote \emph{connected} correlators. These enter the partition function as: 
\begin{align}
	\begin{split}
		Z_{CFT}[A^{(s)}(\mathbf{r})] &= \exp \bigg( \sum_{n=2}^\infty \frac{1}{n!}\int d^3\mathbf{r_1}\sum_{s_1=0}^\infty A^{(s_1)}_{i_1\dots i_{s_1}}(\mathbf{r_1})\ldots \int d^3\mathbf{r_n}\sum_{s_n=0}^\infty A^{(s_n)}_{i_1\dots i_{s_n}}(\mathbf{r_n}) \\
		&\qquad\qquad\qquad\qquad \times \left<J_{(s_1)}^{i_1\dots i_{s_1}}(\mathbf{r_1})\dots J_{(s_n)}^{i_1\dots i_{s_n}}(\mathbf{r_n})\right> \bigg) \ ,
	\end{split} \label{eq:Z_local}
\end{align}
where $A^{(s)}_{i_1\dots i_s}(\mathbf{r})$ are the external gauge-potential sources that couple to the currents $J^{(s)}_{i_1\dots i_s}(\mathbf{r})$. 

Now, a key observation of \cite{Didenko:2012tv,Neiman:2017mel,David:2020ptn} is that the boundary correlators $\langle J^{(s_1)}\dots J^{(s_n)} \rangle$ can be expressed as traced star-products of appropriate twistor functions, which are just the Penrose transforms of the corresponding boundary-bulk propagators. To express this, we first covariantize the boundary coordinates $\mathbf{r}$ and polarization vectors $\boldsymbol{\lambda}$ into embedding-space vectors via \eqref{eq:Poincare}:
\begin{align}
	\ell^\mu = \left(\frac{1+\mathbf{r}^2}{2}, \frac{1-\mathbf{r}^2}{2}, \mathbf{r} \right) \ ; \quad \lambda^\mu = (\boldsymbol{\lambda}\cdot\mathbf{r},-\boldsymbol{\lambda}\cdot\mathbf{r},\boldsymbol{\lambda}) \ .
\end{align}
We then construct a twistor $M^a$ as the square root of the polarization vector $\lambda^\mu$ at $\ell^\mu$, via: 
\begin{align}
	\gamma_{\mu\nu}^{ab}\ell^\mu\lambda^\nu = (\ell M)^a(\ell M^b) \ .
\end{align}
With these definitions, the twistor functions corresponding to the currents $J^{(s)}(\ell;\lambda)$ read \cite{David:2020ptn}:
\begin{align}
	\kappa^{(s)}_{\ell,\lambda}(Y) = \pm \frac{M^{a_1}\dots M^{a_{2s}}}{8\pi} \left(Y_{a_1}\dots Y_{a_{2s}} + (-1)^s\frac{\del^{2s}}{\del Y^{a_1}\dots\del Y^{a_{2s}}}\right) \delta_\ell(Y) \ . \label{eq:kappa_EAdS_type_A}
\end{align}
The star-product formula for correlators then takes the form:
\begin{align}
	\begin{split}
		&\left<J^{(s_1)}(\ell_1;\lambda_1)\dots J^{(s_n)}(\ell_n;\lambda_n)\right> \\
		&\quad = -\frac{N}{4} \Big(\tr_\star\!\big(\kappa^{(s_1)}_{\ell_1,\lambda_1}(Y)\star\ldots\star\kappa^{(s_n)}_{\ell_n,\lambda_n}(Y) \big) + \text{permutations}\Big) \ .
	\end{split} \label{eq:HS_correlators_EAdS_type_A}
\end{align}
The $\pm$ sign ambiguity in \eqref{eq:kappa_EAdS_type_A} is meant to match the one in the 3-point star-product \eqref{eq:3_pt_product}, in which case we obtain the prescribed sign in the $n$-point product \eqref{eq:HS_correlators_EAdS_type_A}. We will return to this sign ambiguity in section \ref{sec:EAdS:partition_function}.

\subsection{Type-A dS/CFT} \label{sec:boundary:type_A_dS}

We now set out to analytically continue the type-A boundary theory \eqref{eq:EAdS_type_A}-\eqref{eq:J_EAdS_type_A} from Euclidean AdS/CFT to dS/CFT. As noted in \cite{Anninos:2011ui}, the 2-point function for the minimal theory, i.e. for even spins, receives the correct sign if the fields $\phi^I,\bar\phi_I$ in \eqref{eq:EAdS_type_A} are taken to be \emph{anti-}commuting, flipping the standard spin-statistics relation. The effect of this on the fundamental correlator \eqref{eq:correlator_EAdS_type_A} is to flip its overall sign (this is nothing but the usual minus sign on a fermionic loop diagram):
\begin{align}
	\big\langle \calO(\mathbf{r_1},\mathbf{r'_1})\dots \calO(\mathbf{r_n},\mathbf{r'_n}) \big\rangle = -\frac{N}{(4\pi)^n} \left(\prod_{p=1}^n \frac{1}{|\mathbf{r_{p+1} - r'_p}|} + \text{permutations}\right) \ . \label{eq:correlator_dS_type_A}
\end{align}
This overall sign propagates to the star-product correlator formula \eqref{eq:HS_correlators_EAdS_type_A}, as:
\begin{align}
	\begin{split}
		&\left<J^{(s_1)}(\ell_1;\lambda_1)\dots J^{(s_n)}(\ell_n;\lambda_n)\right> \\
		&\quad = \frac{N}{4} \Big(\tr_\star\!\big(\kappa^{(s_1)}_{\ell_1,\lambda_1}(Y)\star\ldots\star\kappa^{(s_n)}_{\ell_n,\lambda_n}(Y) \big) + \text{permutations}\Big) \ .
	\end{split} \label{eq:HS_correlators_dS_type_A}
\end{align}
Now, as foreshadowed above (and will be verified in section \ref{sec:boundary:2_point}), the 2-point functions of \emph{odd} spins need an additional sign flip. This can be accomplished by removing the factor of $i^s$ from the definition of the currents \eqref{eq:J_EAdS_type_A}. However, that would make the odd-spin currents imaginary. What we need for consistency is a new reality condition for the vector model's fundamental fields, such that the currents without the factor of $i^s$ would become real.

To accomplish this goal, we note that the color group $U(N)$ of \eqref{eq:EAdS_type_A} is a real section of $GL(N,\bbC)$. We then replace it by a different real section, namely the real $GL(N)$. Instead of the complex-conjugate fields $\phi^I$ and $\bar\phi_I$, we define two \emph{independent, real} sets of fields $\phi^I$ and $\tilde\phi_I$:
\begin{align}
	\bar\phi^I = \phi^I \ ; \quad \overline{\tilde\phi}_I = \tilde\phi_I \ , \label{eq:phi_reality}
\end{align}
which transform respectively in the fundamental and co-fundamental representations of $GL(N)$. The vector model's action otherwise retains the same form:
\begin{align}
	S_{\text{CFT}} = \int d^3\mathbf{r}\,(\boldsymbol{\nabla}\tilde\phi_I)\cdot(\boldsymbol{\nabla}\phi^I) \ , \label{eq:dS_type_A}
\end{align}
where we still take $\phi^I$ and $\tilde\phi_I$ to be anti-commuting, so that the correlators of bilocals $\calO(\mathbf{r},\mathbf{r'}) \equiv \tilde\phi_I(\mathbf{r'})\phi^I(\mathbf{r})$ continue to take the dS/CFT form \eqref{eq:correlator_dS_type_A}. By design, the real single-trace currents $J^{(s)}$ now read:
\begin{align}
	J^{(s)}(\mathbf{r};\boldsymbol{\lambda}) = \sum_{k=0}^s (-1)^k \binom{2s}{2k} (\boldsymbol{\lambda\cdot\nabla})^k (\boldsymbol{\lambda\cdot\nabla'})^{s-k} \, \tilde\phi_I(\mathbf{r'})\phi^I(\mathbf{r}) \Big|_{\mathbf{r'=r}} \ , \label{eq:J_dS_type_A}
\end{align}
without the $i^s$ factors from \eqref{eq:J_EAdS_type_A}. This completes our definition of the boundary dual for non-minimal type-A HS Gravity. For the star-product correlator formula \eqref{eq:HS_correlators_dS_type_A} to hold, we need to adjust the ``boundary-twistor propagators'' \eqref{eq:kappa_EAdS_type_A} by the same factor of $i^s$, as:
\begin{align}
	\kappa^{(s)}_{\ell,\lambda}(Y) = \pm \frac{M^{a_1}\dots M^{a_{2s}}}{8\pi i^s} \left(Y_{a_1}\dots Y_{a_{2s}} + (-1)^s\frac{\del^{2s}}{\del Y^{a_1}\dots\del Y^{a_{2s}}}\right) \delta_\ell(Y) \ . \label{eq:kappa_dS_type_A}
\end{align}
The boundary theory \eqref{eq:dS_type_A}-\eqref{eq:J_dS_type_A} can be restricted to even-spin currents, recovering the minimal model of \cite{Anninos:2011ui}. To do this, we unify $\phi^\scri\equiv (\phi^I,\tilde\phi_I)$ into a single multiplet of an enlarged color group $Sp(2N)\supset GL(N)$, which preserves the symplectic form:
\begin{align}
	\Omega_{\scri\calJ} = \frac{1}{2}\begin{pmatrix} 0 & -\delta_I{}^J \\ \delta^I{}_J & 0 \end{pmatrix} \ . \label{eq:Omega}
\end{align}
The action \eqref{eq:dS_type_A} can be rewritten as:
\begin{align}
	S_{\text{CFT}} = \int d^3\mathbf{r}\,\Omega_{\scri\calJ}(\boldsymbol{\nabla}\phi^\scri)\cdot(\boldsymbol{\nabla}\phi^\calJ) \ . \label{eq:dS_type_A_minimal}
\end{align}
Out of the currents \eqref{eq:J_dS_type_A}, only those with even spins $s$ are invariant under the entire $Sp(2N)$. They can be written as:
\begin{align}
	J^{(s)}(\mathbf{r};\boldsymbol{\lambda}) = \sum_{k=0}^s (-1)^k \binom{2s}{2k} (\boldsymbol{\lambda\cdot\nabla})^k (\boldsymbol{\lambda\cdot\nabla'})^{s-k} \,\Omega_{\scri\calJ}\phi^\scri(\mathbf{r'})\phi^\calJ(\mathbf{r}) \Big|_{\mathbf{r'=r}} \ . \label{eq:J_dS_type_A_minimal}
\end{align}
Let us end with a general comment on boundary theories in dS/CFT. Our change of the color group from the compact $U(N)$ to the non-compact $GL(N)$ follows the same trend as the spin-statistics flip from \cite{Anninos:2011ui}: both mean that \emph{the Euclidean boundary CFT is not the Wick rotation of a unitary Lorentzian CFT}. We wish to emphasize that this is not a problem: unlike in Euclidean AdS/CFT, the boundary theory is \emph{never meant} to be made Lorentzian: the unitary Lorentzian theory we want is in the de Sitter bulk. See \cite{Thavanesan:2025ibm} for a nice discussion of this point in the case of Yang-Mills-like boundary theories.

\subsection{Type-B Euclidean AdS/CFT} \label{sec:boundary:type_B_EAdS}

We now turn to the type-B theory. In Euclidean AdS/CFT, its boundary dual is again a free $U(N)$ vector model, but with spinor fields $\psi^I,\bar\psi_I$ in place of the scalar fields $\phi^I,\bar\phi_I$. More precisely, using $SU(2)$ spinor indices $(\alpha,\beta,\dots)$, our fundamental fields are $\psi^{\alpha I}$, with complex conjugates $\bar\psi_{\alpha I}$. Since for the moment we're still in AdS/CFT, these fields obey spin-statistics, i.e. are anti-commuting. The action reads:
\begin{align}
	S_{\text{CFT}} = \int d^3\mathbf{r}\,\bar\psi_I(\boldsymbol{\sigma\cdot\nabla})\psi^I \ , \label{eq:EAdS_CFT_type_B}
\end{align}
where $\boldsymbol{\sigma}$ are the Pauli matrices. Note the action \eqref{eq:EAdS_CFT_type_B} is, up to integration by parts, \emph{imaginary}. This is an artifact of analytically continuing fermions from Lorentzian to Euclidean: the imaginary action \eqref{eq:EAdS_CFT_type_B} is the one whose correlators agree with those of Lorentzian AdS/CFT (on a constant-time slice, where the two can be compared). The single-trace primary operators of the CFT \eqref{eq:EAdS_CFT_type_B} are again a tower of spin-$s$ currents for all integer $s$, constructed from the bilocal single-trace operator $\calO^\beta{}_\alpha(\mathbf{r},\mathbf{r'}) \equiv \bar\psi_{\alpha I}(\mathbf{r'})\psi^{\beta I}(\mathbf{r})$. Unlike the type-A case, this time the spin-0 ``current'' stands apart from the true, nonzero-spin currents. Specifically, we have:
\begin{align}
	\begin{split}
		J^{(0)}(\mathbf{r}) &= \bar\psi_I(\mathbf{r})\psi^I(\mathbf{r}) \ ; \\
	    J^{(s>0)}(\mathbf{r};\boldsymbol{\lambda}) &= \frac{i^{s-1}}{2^s} \sum_{k=0}^{s-1} (-1)^k \binom{2s}{2k+1} (\boldsymbol{\lambda\cdot\nabla})^k (\boldsymbol{\lambda\cdot\nabla'})^{s-1-k} \, 
	       \bar\psi_I(\mathbf{r'})(\boldsymbol{\lambda\cdot\sigma})\psi^I(\mathbf{r}) \Big|_{\mathbf{r'=r}} \ , 
	\end{split} \label{eq:J_EAdS_type_B}
\end{align}
where the factors of $i$ are such that the currents \eqref{eq:J_EAdS_type_B} are all real.

The $n$-point correlators of the type-B theory can again all be assembled from the correlator of $n$ bilocals. These are given by a formula similar to \eqref{eq:correlator_EAdS_type_A}, but with fermion propagators instead of the scalar ones, and with an overall minus sign due to the fermion loop. What's important for our purposes is that the $n$-point correlators can be computed \cite{Didenko:2013bj,Lang:2024dkt} by the exact same star-product formula \eqref{eq:HS_correlators_dS_type_A} as in the type-A theory. The only difference is in the choice of ``boundary-twistor propagators'' $\kappa^{(s)}_{\ell,\lambda}(Y)$. For zero and nonzero spins respectively, these read:
\begin{align}
  \begin{split}
     \kappa^{(0)}_\ell(Y) &= \pm\frac{i}{8\pi(\ell\cdot\ell_0)}Y\ell_0\frac{\partial}{\partial Y}\delta_\ell(Y) \ ; \\
     \kappa^{(s>0)}_{\ell,\lambda}(Y) &= \pm \frac{iM^{a_1}\dots M^{a_{2s}}}{8\pi} \left(Y_{a_1}\dots Y_{a_{2s}} - (-1)^s\frac{\del^{2s}}{\del Y^{a_1}\dots\del Y^{a_{2s}}}\right) \delta_\ell(Y) \ . 
   \end{split} \label{eq:kappa_EAdS_type_B} 
\end{align}
Here, in the formula for $\kappa^{(0)}_\ell(Y)$, we used an auxiliary lightlike direction $\ell_0^\mu\nsim \ell^\mu$ in embedding space. The final result does not depend on the choice of $\ell_0^\mu$. 

\subsection{Type-B dS/CFT} \label{sec:boundary:type_B_dS}

We now present the analytic continuation of the type-B theory \eqref{eq:EAdS_CFT_type_B}-\eqref{eq:J_EAdS_type_B} from Euclidean AdS/CFT to dS/CFT. We repeat the two steps we performed for the type-A theory in section \ref{sec:boundary:type_A_dS}:  flip the spin-statistics (as suggested in \cite{Hertog:2017ymy}), then flip the sign of the odd-spin 2-point functions by changing the color group from $U(N)$ to $GL(N)$. In a third step specific to type-B, we will further flip the sign of \emph{all} the 2-point functions, by multiplying the boundary action \eqref{eq:EAdS_CFT_type_B} by $i$. Since the original action \eqref{eq:EAdS_CFT_type_B} is imaginary, this makes the dS/CFT boundary action \emph{real}. We were motivated to take this step by the fact that the type-B boundary-bulk propagators in \cite{Lang:2024dkt} carry an extra factor of $i$. At the 2-point level, this step has the same effect as the spin-statistics flip, so we could equally well skip them both. However, that would spoil the global-maximum property of the partition function in section \ref{sec:EAdS} below.

Let's now proceed to explicit definitions. The fundamental boundary fields are two multiplets of \emph{commuting} $SU(2)$ spinors $\psi^I(\mathbf{r})$ and $\tilde\psi_I(\mathbf{r})$, which transform respectively in the fundamental and co-fundamental representations of the color group $GL(N)$. As in the type-A case of section \ref{sec:boundary:type_A_dS}, $\psi^I$ and $\tilde\psi_I$ are not complex-conjugate to each other, but rather each of them should satisfy a reality condition. This is complicated by the fact that the complex conjugation $\chi_\alpha\to\bar\chi_\alpha$ of $SU(2)$ spinors is anti-idempotent $\bar{\bar\chi}_\alpha = -\chi_\alpha$. Thus, to define a consistent reality condition, we introduce an anti-idempotent matrix (i.e. a complex structure) $\calJ^I{}_J$ in color space:
\begin{align}
	\calJ^I{}_J\calJ^J{}_K = -\delta^I_K \ .
\end{align}
For such a matrix to exist, $N$ must be even. The reality conditions then read:
\begin{align}
	\bar\psi^I = \calJ^I{}_J\,\psi^J \ ; \quad \overline{\tilde\psi}_I = -\calJ^J{}_I\,\tilde\psi_J \ , \label{eq:psi_reality}
\end{align}
such that $\tilde\psi_I\psi^I$ is real. We then define the boundary action as:
\begin{align}
	S_{\text{CFT}} = i\int d^3\mathbf{r}\,\tilde\psi_I(\boldsymbol{\sigma\cdot\nabla})\psi^I \ . \label{eq:dS_CFT_type_B}
\end{align}
As mentioned above, unlike in Euclidean AdS/CFT \eqref{eq:EAdS_CFT_type_B}, this action is real. The real spin-$s$ currents read:
\begin{align}
  \begin{split}
	J^{(0)}(\mathbf{r}) &= \tilde\psi_I(\mathbf{r})\psi^I(\mathbf{r}) \ ; \\
	J^{(s>0)}(\mathbf{r};\boldsymbol{\lambda}) &= -\frac{i}{2^s} \sum_{k=0}^{s-1} (-1)^k \binom{2s}{2k+1} (\boldsymbol{\lambda\cdot\nabla})^k (\boldsymbol{\lambda\cdot\nabla'})^{s-1-k} \, 
	\tilde\psi_I(\mathbf{r'})(\boldsymbol{\lambda\cdot\sigma})\psi^I(\mathbf{r}) \Big|_{\mathbf{r'=r}} \ .
  \end{split} \label{eq:J_dS_type_B}
\end{align}
As in section \ref{sec:boundary:type_A_dS}, these formulas differ from the Euclidean AdS/CFT ones \eqref{eq:J_EAdS_type_B} by a factor of $1/i^s$. Note that the action \eqref{eq:dS_CFT_type_B} and single-trace currents \eqref{eq:J_dS_type_B} are both invariant under the full $GL(N)$ color group, even though the matrix $\calJ^I{}_J$ and the reality condition \eqref{eq:psi_reality} are not.

The HS-algebraic formula for the correlators is the same as in the type-A case \eqref{eq:HS_correlators_dS_type_A}, with opposite sign from the Euclidean AdS/CFT formula \eqref{eq:HS_correlators_EAdS_type_A} due to the flipped statistics of the boundary fields. For the formula \eqref{eq:HS_correlators_dS_type_A} to hold, we must adjust the ``boundary-twistor propagators'' from Euclidean AdS/CFT \eqref{eq:kappa_EAdS_type_B} by a factor of $1/i^{s+1}$. This consists of a factor of $1/i^s$ for the difference between the current formulas \eqref{eq:J_EAdS_type_B} and \eqref{eq:J_dS_type_B}, and another factor of $1/i$ for the difference between the propagators arising from the actions \eqref{eq:EAdS_CFT_type_B} and \eqref{eq:dS_CFT_type_B}. Thus, the dS/CFT ``boundary-twistor propagators'' read:
\begin{align}
	\begin{split}
		\kappa^{(0)}_\ell(Y) &= \pm\frac{1}{8\pi(\ell\cdot\ell_0)}Y\ell_0\frac{\partial}{\partial Y}\delta_\ell(Y) \ ; \\
		\kappa^{(s>0)}_{\ell,\lambda}(Y) &= \pm \frac{M^{a_1}\dots M^{a_{2s}}}{8\pi i^s} \left(Y_{a_1}\dots Y_{a_{2s}} - (-1)^s\frac{\del^{2s}}{\del Y^{a_1}\dots\del Y^{a_{2s}}}\right) \delta_\ell(Y) \ . 
	\end{split} \label{eq:kappa_dS_type_B} 
\end{align}
The theory \eqref{eq:dS_CFT_type_B}-\eqref{eq:J_dS_type_B} can be restricted to even-spin currents, forming the boundary dual of \emph{minimal} type-B HS Gravity in de Sitter. As in the type-A case of section \ref{sec:boundary:type_A_dS}, we unify $\psi^\scri\equiv (\psi^I,\tilde\psi_I)$ into a single multiplet of an enlarged color group $Sp(2N)\supset GL(N)$, which preserves the symplectic form \eqref{eq:Omega}. The action \eqref{eq:dS_CFT_type_B} is then rewritten as:
\begin{align}
	S_{\text{CFT}} = i\int d^3\mathbf{r}\,\Omega_{\scri\calJ}\,\psi^\scri(\boldsymbol{\sigma\cdot\nabla})\psi^\calJ \ . \label{eq:dS_type_B_minimal}
\end{align}
Out of the currents \eqref{eq:J_dS_type_B}, only those with even spins $s$ are invariant under the entire $Sp(2N)$. They can be written as:
\begin{align}
	J^{(0)}(\mathbf{r}) &= \Omega_{\scri\calJ}\,\psi^\scri(\mathbf{r})\psi^\calJ(\mathbf{r}) \ ; \label{eq:J_dS_type_B_minimal} \\
    J^{(s>0)}(\mathbf{r};\boldsymbol{\lambda}) &= -\frac{i}{2^s} \sum_{k=0}^{s-1} (-1)^k \binom{2s}{2k+1} (\boldsymbol{\lambda\cdot\nabla})^k (\boldsymbol{\lambda\cdot\nabla'})^{s-1-k} \,\Omega_{\scri\calJ}\,\psi^\scri(\mathbf{r'})(\boldsymbol{\lambda\cdot\sigma})\psi^\calJ(\mathbf{r}) \Big|_{\mathbf{r'=r}} \ . \nonumber
\end{align}
The reality condition \eqref{eq:psi_reality} takes the form:
\begin{align}
	\bar\psi^\scri = \calJ^\scri{}_\calJ\,\psi^\calJ \ ,
\end{align}
with the $2N\times 2N$ color matrix $\calJ^\scri{}_\calJ$:
\begin{align}
	\calJ^\scri{}_\calJ = \begin{pmatrix} \calJ^I{}_J & 0 \\ 0 & -\calJ^J{}_I \end{pmatrix} \ .
\end{align}
In the minimal theory \eqref{eq:dS_type_B_minimal}-\eqref{eq:J_dS_type_B_minimal}, this can be replaced by any complex structure $\calJ^\scri{}_\calJ$ that is compatible with the symplectic form $\Omega_{\scri\calJ}$:
\begin{align}
	\calJ^\scri{}_\calJ\calJ^\calJ{}_\calK = -\delta^\scri_\calK \ ; \quad \calJ^\calK{}_\scri\,\Omega_{\calJ\calK} = \calJ^\calK{}_\calJ\,\Omega_{\scri\calK} \ . \label{eq:J_Omega}
\end{align}
We can then drop the requirement that $N$ is even.

We end with a comment on supersymmetry. In Euclidean AdS/CFT, the type-A and type-B theories of sections \ref{sec:boundary:type_A_EAdS},\ref{sec:boundary:type_B_EAdS} can easily be unified into a supersymmetric theory, by adding together the actions \eqref{eq:EAdS_type_A},\eqref{eq:EAdS_CFT_type_B}, and including fermionic currents of the schematic form $\bar\phi_I\psi^I$ and $\bar\psi_I\phi^I$. For the dS/CFT boundary theories of sections \ref{sec:boundary:type_A_dS},\ref{sec:boundary:type_B_dS}, the situation is different. Of course, one can still add together the actions \eqref{eq:dS_type_A} and \eqref{eq:dS_CFT_type_B}, producing a union of the type-A and type-B theories. However, one can't extend this into a supersymmetric theory, because there is no way to introduce fermionic currents. This is due to the difference between the trivial reality condition \eqref{eq:phi_reality} on the fundamental fields $\tilde\phi_I,\phi^I$ in the type-A sector, and the non-trivial reality condition \eqref{eq:psi_reality} on $\tilde\psi_I,\psi^I$ in the type-B sector. Due to this difference, one cannot construct combinations of the form $\tilde\phi_I\psi^I$ or $\tilde\psi_I\phi^I$ that would be both real and invariant under the full $GL(N)$ or $Sp(2N)$ color group.

\subsection{Boundary-bulk propagators and the signs of 2-point functions} \label{sec:boundary:2_point}

Let us now verify that the type-A and type-B boundary theories of sections \ref{sec:boundary:type_A_dS},\ref{sec:boundary:type_B_dS} have the correct positivity properties for dS/CFT at the 2-point level. First, let us write the boundary-bulk propagators $C^{(s)}_{\ell,\lambda}(x;Y) \equiv i\kappa^{(s)}_{\ell,\lambda}(Y)\star\delta_x(Y)$ that correspond, via the Penrose transform \eqref{eq:Penrose}, to the ``boundary-twistor propagators'' \eqref{eq:kappa_dS_type_A},\eqref{eq:kappa_dS_type_B}. These can be read off from the results of \cite{David:2020ptn,Lang:2024dkt}, adjusting by appropriate powers of $i$. The results read, for type-A:
\begin{align}
  C^{(s)}_{\ell,\lambda}(x;Y) = \pm \frac{i^{s+1}}{4\pi}\frac{(M\ell P_x Y)^{2s} + (M\ell P_{-x} Y)^{2s}}{(\ell\cdot x)^{2s+1}} \exp\frac{iY\ell xY}{2(\ell\cdot x)} \ , \label{eq:type_A_propagator}
\end{align}
and for type-B:
\begin{align}
	\begin{split}
  	  C^{(0)}_\ell(x;Y) &= \frac{\mp 1}{2\pi(\ell\cdot x)^2}\left(1+\frac{iY\ell xY}{2(\ell\cdot x)}\right)\exp\frac{iY\ell xY}{2(\ell\cdot x)} \ ; \\
	  C^{(s>0)}_{\ell,\lambda}(x;Y) &= \pm \frac{i^{s-1}}{4\pi}\frac{(M\ell P_x Y)^{2s} - (M\ell P_{-x} Y)^{2s}}{(\ell\cdot x)^{2s+1}}\exp\frac{iY\ell xY}{2(\ell\cdot x)} \ . 
	\end{split} \label{eq:type_B_propagator}
\end{align}
The propagator master fields \eqref{eq:type_A_propagator}-\eqref{eq:type_B_propagator} decompose neatly into spin-$s$ field strengths \eqref{eq:packaging} and their towers of derivatives \eqref{eq:unfolding}. Specifically, the towers of derivatives are encoded in the functions of $(iY\ell xY)/(2\ell\cdot x)$, while the field strengths themselves are encoded in their $(\ell\cdot x)$- and $(M\ell P_{\pm x} Y)$-dependent prefactors.

Now, let's consider the reality properties of our boundary-bulk propagators under the complex conjugation \eqref{eq:new_conjugation}. These can be read off directly from \eqref{eq:new_conjugation} for the bulk master-fields \eqref{eq:type_A_propagator}-\eqref{eq:type_B_propagator}, or via \eqref{eq:star_identities} for the twistor functions \eqref{eq:kappa_dS_type_A},\eqref{eq:kappa_dS_type_B}. When performing the complex conjugations, we assume that the boundary point $\ell^\mu$ is real. The null polarization vector $\lambda^\mu$ is necessarily complex. We also allow the bulk point $x^\mu$ to be complex, so as to cover both real $dS_4$ points and imaginary $EAdS_4^\pm$ points. With this understanding, the reality properties read, for both type-A and type-B:
\begin{align}
	\kappa^{(s)*}_{\ell,\lambda}(Y) = \kappa^{(s)}_{\ell,\bar\lambda}(Y) \ ; \quad C^{(s)*}_{\ell,\lambda}(x;Y) = C^{(s)}_{\ell,\bar\lambda}(-\bar x;Y) \ . \label{eq:propagators_reality}
\end{align}
Recall that the appearance of a complex polarization vector $\lambda^\mu$ is an artefact of our packaging the indices of the boundary currents in \eqref{eq:J_packaging}. We can undo it by unpacking these indices on the propagators:
\begin{align}
	C^{(s)}_{\ell,\lambda}(x;Y) = C^{(s)\mu_1\dots\mu_s}_\ell(x;Y)\,\lambda^{\mu_1}\dots\lambda^{\mu_s} \ ,
\end{align}
and integrating against real boundary sources $A^{(s)}_{\mu_1\dots\mu_s}(\ell)$ as in \eqref{eq:Z_local} (with an $N$-dependent overall factor for later convenience):
\begin{align}
	C(x_E;Y) &\equiv \sqrt{\frac{N}{4}}\int d^3\ell \sum_s A^{(s)}_{\mu_1\dots\mu_s}(\ell)\,C^{(s)\mu_1\dots\mu_s}_\ell(x_E;Y) \ ; \label{eq:integrated_C} \\
	F(Y) &\equiv -iC(x_E;Y)\star\delta_{x_E}(Y) \ . \label{eq:integrated_F}
\end{align}
Here, $x_E$ is an arbitrary origin point in $EAdS_4^+$, as in \eqref{eq:x_E}. The reality property \eqref{eq:propagators_reality} now becomes:
\begin{align}
	F^*(Y) = F(Y) \ ; \quad C^*(x_E;Y) = C(x_E;Y) \ . \label{eq:integrated_reality}
\end{align}
which we recognize as the condition \eqref{eq:reality_star_EAdS} that \emph{the bulk fields are real at imaginary $x^\mu$, i.e. in Euclidean AdS} (recalling that, for odd spins, the notion of reality here is flipped from the standard one). 

We pause to note the order of operations we chose in \eqref{eq:integrated_C}-\eqref{eq:integrated_F}, as well our choice of the imaginary $x_E\in EAdS_4^+$ as the bulk point to consider. A priori, we could define $F(Y)$ by directly integrating the boundary-twistor propagators $\kappa^{(s)}$; and, if we do refer to a bulk point, then the $x_E$ in \eqref{eq:integrated_C}-\eqref{eq:integrated_F} can be replaced by any other point $x$, in particular a real one $x\in dS_4$. However, this ignores the ambiguity in integrating the poles of the propagators \eqref{eq:type_A_propagator}-\eqref{eq:type_B_propagator}. The specific construction \eqref{eq:integrated_C}-\eqref{eq:integrated_F} effectively imposes an $i\varepsilon$ prescription on the propagators, which results in positive-frequency bulk fields $C(x;Y)$, i.e. ones that are regular on $EAdS_4^+$, in keeping with the dS/CFT paradigm of \cite{Maldacena:2002vr,Anninos:2011ui} (we will venture beyond this paradigm in section \ref{sec:elliptic_dS}).

A good basis for the positive-frequency fields \eqref{eq:integrated_C}-\eqref{eq:integrated_F} is the basis \eqref{eq:C_Psi}-\eqref{eq:F_Psi} that we constructed for particle wavefunctions in section \ref{sec:Hermitian:basis}. While those wavefunctions were complex, we now have the reality condition \eqref{eq:integrated_reality}. For the coefficients in \eqref{eq:C_Psi}-\eqref{eq:F_Psi}, this implies:
\begin{align}
	\bar\Psi^{(m,n)}_{a_1\dots a_m b_1\dots b_n} = (-i)^{m+n}\,\Psi^{(m,n)}_{a_1\dots a_m b_1\dots b_n} \ . \label{eq:Psi_reality}
\end{align}
Note that this reality condition automatically imposes a restriction to integer spins, i.e. to even $m+n$. 

Finally, we are ready to verify the sign of the dS/CFT 2-point functions. The desired property is for the $n=2$ (Gaussian) part of the partition function \eqref{eq:Z_local} to be the exponent of a \emph{negative-definite} quadratic form, so that it has a maximum at vanishing sources. Recalling the star-product formula \eqref{eq:HS_correlators_dS_type_A} for dS/CFT correlators, this becomes a condition on the integrated boundary-bulk propagator \eqref{eq:integrated_C}-\eqref{eq:integrated_F}:
\begin{align}
	\tr_\star\big(F(Y)\star F(Y)\big) < 0 \ . \label{eq:2_point_condition}
\end{align}
Recalling the reality property \eqref{eq:integrated_reality} of the integrated propagators, we recognize the LHS of \eqref{eq:2_point_condition} as minus the RHS of the Hermitian norm formula \eqref{eq:Psi^2_F}, which we already established to be positive-definite. The desired property \eqref{eq:2_point_condition} is thus satisfied. This completes our check that the boundary theories of sections \ref{sec:boundary:type_A_dS},\ref{sec:boundary:type_B_dS} give 2-point functions with the correct signs for dS/CFT.

Note that the sign ambiguities in the propagators \eqref{eq:kappa_dS_type_A},\eqref{eq:kappa_dS_type_B},\eqref{eq:type_A_propagator}-\eqref{eq:type_B_propagator} don't play any role in the above positivity analysis. These sign ambiguities are inherited from the one in the 3-point product \eqref{eq:3_pt_product}, and do not affect the 2-point-level discussion.

\section{Type-A and type-B identified in the Euclidean} \label{sec:EAdS}

In this section, we go beyond section \ref{sec:boundary}'s boundary correlators, and consider the dS/CFT partition function at finite sources. As observed in \cite{David:2020ptn}, at this level the equivalence \eqref{eq:HS_correlators_dS_type_A} between HS star-products and the boundary CFT breaks down, and the HS-algebraic expression becomes better-behaved than the original CFT partition function. In section \ref{sec:EAdS:partition_function}, we review these observations. In section \ref{sec:EAdS:spin_1}, we demonstrate that the good behavior of the HS-algebraic partition function extends to odd spins, by showing that it has a global maximum at vanishing sources not only for the lowest spin-0 harmonic, but also for the lowest spin-1 harmonic. In section \ref{sec:EAdS:Legendre}, we observe that the HS-algebraic partition function is self-dual under the Legendre transform. We then use this to advance the non-standard notion, first proposed in \cite{Lang:2024dkt}, that type-A and type-B can be viewed as the same theory with opposite boundary conditions.

\subsection{HS-algebraic vs. boundary-local partition function} \label{sec:EAdS:partition_function}

We start by recalling the relation \eqref{eq:Z_local}, which assembles the partition function from the boundary correlators. As we already did at the 2-point level in section \ref{sec:boundary:2_point}, we can use the HS-algebraic formula \eqref{eq:HS_correlators_dS_type_A} for the correlators to express this as:
\begin{align}
  \begin{split}
	Z_{HS}[F(Y)] &= \exp\tr_\star\left(\frac{N}{4} \sum_{n=2}^\infty \frac{1}{n}\left(\frac{4}{N}\right)^{n/2}\underbrace{F(Y)\star\ldots\star F(Y)}_{n\text{ factors}} \right) \\
	  &= \exp\tr_\star\left(-\frac{N}{4}\ln_\star\left[1-\sqrt{\frac{4}{N}}\,F(Y)\right] - \sqrt{\frac{N}{4}}\,F(Y) \right) \ ,
  \end{split} \label{eq:Z_HS}
\end{align}
where the twistor function $F(Y)$ is the Penrose transform \eqref{eq:integrated_F} of the superposition \eqref{eq:integrated_C} of boundary-bulk propagators, with coefficients given by the boundary sources $A^{(s)}_{\mu_1\dots\mu_s}(\ell)$. The factors of $1/n$ in the first line of \eqref{eq:Z_HS} arise, as usual for a 1-loop Feynman diagram, from multiplying the $1/n!$ factors in \eqref{eq:Z_local} by the $(n-1)!$ permutations in \eqref{eq:HS_correlators_dS_type_A}. The factors of $\sqrt{4/N}$ cancel the ones introduced in \eqref{eq:integrated_F}. The notation $\ln_\star[\dots]$ in the second line of \eqref{eq:Z_HS} denotes the usual logarithm $\ln[\dots]$, with every product in the Taylor expansion replaced by a star-product.

The surprising observation of \cite{Lang:2024dkt} is that beyond the 2-point level, the HS-algebraic partition function \eqref{eq:Z_HS} does \emph{not} agree with the original CFT partition function \eqref{eq:Z_local}, despite the agreement \eqref{eq:HS_correlators_dS_type_A} of the correlators. In other words, when the correlators are integrated against boundary sources, the superposition principle fails. As discussed in \cite{Lang:2024dkt}, this failure/disagreement is related to, but distinct from, the need to correct \eqref{eq:Z_local} by contact terms. In particular, the disagreement occurs already for spin-0 sources, where contact terms do not arise. 

Of the explanations for the disagreement given in \cite{Lang:2024dkt}, the one closest to our present discussion is in terms of \emph{spontaneous $\bbZ_2$ symmetry breaking}. Specifically, consider the standard complex conjugation $f(Y)\to f^\dagger(Y)$ from \eqref{eq:old_conjugation}, and the ``spin-parity'' operation $Y^a\to iY^a$. Both of these are $\bbZ_2$ symmetries of the HS algebra, which reverse the order of the star-product. When combined, they produce the ``new'' complex conjugation $f(Y)\to f^*(Y)$ from \eqref{eq:new_conjugation}, which preserves the order of the star-product. Now, recall the sign ambiguity in the 3-point product \eqref{eq:3_pt_product}, which continues into all our propagators \eqref{eq:kappa_dS_type_A},\eqref{eq:kappa_dS_type_B},\eqref{eq:type_A_propagator}-\eqref{eq:type_B_propagator}. Whichever sign we choose in \eqref{eq:3_pt_product}, the symmetries $f(Y)\to f^\dagger(Y)$ and $Y^a\to iY^a$ (but not their combination $f(Y)\to f^*(Y)$) end up violated; only the existence of the second sign choice recovers the symmetries. Thus, the HS star-products that compute the correlators \eqref{eq:HS_correlators_dS_type_A}, and from there the standard partition function \eqref{eq:Z_local}, break the $\bbZ_2$ symmetries. On the other hand, the HS star-products that directly compute the partition function at finite sources via \eqref{eq:Z_HS} preserve these symmetries. This is how the two partition functions end up disagreeing.

The other observation in \cite{Lang:2024dkt} was that the HS-algebraic partition function \eqref{eq:Z_HS} is \emph{better-behaved} than the standard one \eqref{eq:Z_local}, leading to the proposal that it's the correct one to use for dS/CFT. Specifically, the finding is that, at least for a constant spin-0 source on the $S_3$ boundary, the HS-algebraic partition function \eqref{eq:Z_HS} has a \emph{global maximum at vanishing source}. This is in contrast to the standard CFT partition function \eqref{eq:Z_local}, where the maximum is only local \cite{Anninos:2012ft}. To describe this result explicitly, we return to our basis \eqref{eq:C_Psi}-\eqref{eq:F_Psi}, which precisely corresponds to bulk fields, or boundary sources, classified by $S_3$ spherical harmonics. In this basis, a constant spin-0 source on the $S_3$ boundary is described by turning on the coefficient $\Psi^{(0,0)}\equiv \left(\sqrt{N/4}\right)c$, with all the higher coefficients $\Psi^{(m,n)}$ vanishing. The corresponding twistor function $F(Y)$, and master field $C(x_E;Y)$ at the Euclidean origin point $x_E$, read simply:
\begin{align}
	C(x_E;Y) = \sqrt{\frac{N}{4}}\,c \ ; \quad F(Y) = -\sqrt{\frac{N}{4}}\,ic\,\delta_{x_E}(Y) \ . \label{eq:C_F_spin_0}
\end{align}
Let us now plug this into the HS-algebraic partition function \eqref{eq:Z_HS}. The property $\delta_{x_E}\star\delta_{x_E}=1$ from \eqref{eq:delta_delta} means that the even vs. odd powers $n$ in \eqref{eq:Z_HS} can be collected as:
\begin{align}
	Z_{HS}(c) = \exp\left(-\frac{N}{8}\ln(1+c^2) + \frac{iN}{4}(\arctan c - c)\tr_\star\delta_{x_E}(Y)\right) \ .
\end{align}
As argued in \cite{David:2020ptn}, the ill-defined $\tr_\star\delta_{x_E}(Y)$ should be assigned the value zero, as this is the only value that respects the $Y\to iY$ symmetry. We are then left with only the even powers of the source $c$, and the partition function becomes simply:
\begin{align}
	Z_{HS}(c) = \frac{1}{(1+c^2)^{N/8}} \ , \label{eq:Z_spin_0}
\end{align}
which clearly has a global maximum at $c=0$. 

For more general sources composed of finitely many harmonics from the basis \eqref{eq:C_Psi}-\eqref{eq:F_Psi}, the computation of the partition function \eqref{eq:Z_HS} follows a similar pattern. At even orders $n$, the product $F(Y)\star\ldots\star F(Y)$ is a polynomial in $Y$, from which the trace picks out the zeroth power $\tr_\star 1 = 1$. At odd orders $n$, the product $F(Y)\star\ldots\star F(Y)$ contains a factor of $\delta_{x_E}(Y)$, acted on by a polynomial in $Y$ and $\del/\del Y$. As argued in \cite{David:2020ptn}, we take the trace of all such expressions to vanish: the trace of $\delta_{x_E}$ itself vanishes due to the $Y\to iY$ discrete symmetry, while the traces of all other monomials vanish by the $SO(4)$ symmetry of rotations around $x_E$.

\subsection{Global maximum for spin-1} \label{sec:EAdS:spin_1}

In section \ref{sec:boundary}, we extended Higher-Spin dS/CFT to include odd spins, by constructing reality conditions that ensure the correct sign of the 2-point functions. As discussed in section \ref{sec:boundary:2_point}, this simply means that the partition function has a \emph{local maximum} at vanishing sources, for odd spins as well as even ones. Let us now demonstrate that the stronger property of a \emph{global maximum} also extends from the simplest even-spin source mode \eqref{eq:C_F_spin_0}-\eqref{eq:Z_spin_0} to the simplest odd-spin source mode, i.e. to a spin-1 source in the lowest spherical harmonic on $S_3$. Such a source can come in either a right-handed or left-handed version. Without loss of generality, we consider the left-handed mode, where the only non-vanishing coefficient in \eqref{eq:C_Psi}-\eqref{eq:F_Psi} is $\Psi^{(0,2)}_{ab}\equiv \left(\sqrt{N/4}\right)2cM_{ab}$. Here, $c$ is again a scalar parameterizing the source's magnitude, while the polarization $M_{ab}$ is a normalized symmetric matrix on the left-handed spinor space $P(-x_E)$:
\begin{align}
	M_{ab} = M_{ba} \ ; \quad P^a{}_b(x_E)M_{ac} = 0 \ ; \quad \bar M_{ab} = -M_{ab} \ ; \quad M_{ab}M^{ab} = -\frac{1}{2} \ ,
\end{align}
with the reality condition following from \eqref{eq:Psi_reality}. In our component conventions \eqref{eq:U_bar}-\eqref{eq:I_gamma},\eqref{eq:x_E}, a valid example is given by $M_{ab} = \diag(0,0,i/2,i/2)$. The corresponding master field \eqref{eq:C_Psi} and twistor function \eqref{eq:F_Psi} read:
\begin{align}
	C(x_E;Y) = \sqrt{\frac{N}{4}}\,cM_{ab}Y^aY^b \ ; \quad F(Y) = -\sqrt{\frac{N}{4}}\,icM_{ab}Y^aY^b\,\delta_{x_E}(Y) \ . \label{eq:C_F_spin_1}
\end{align}
We now plug this into the HS-algebraic partition function \eqref{eq:Z_HS}:
\begin{align}
	Z_{HS}(c) = \exp\left(\frac{N}{4}\sum_{n=2}^\infty \frac{(-ic)^n}{n} \tr_\star\Big(M_{ab}Y^aY^b\,\delta_{x_E}(Y)\Big)_\star^n\right) \ ,
\end{align}
where $\big(f(Y)\big)_\star^n$ denotes raising $f(Y)$ to the $n$'th power using the star-product.
 
At this point, the fact that we chose a left-handed mode slightly simplifies the calculation. This is because the polynomial factors such as $M_{ab}Y^aY^b$ depend only on the left-handed spinor $y_{(-x_E)}$, while the delta-function $\delta_{x_E}(Y)$ depends only on the right-handed $y_{(x_E)}$. As a result, the star-product between these two types of factors is just the ordinary product of functions. As with the scalar source in section \ref{sec:EAdS:partition_function}, the even and odd powers $n$ combine separately via $\delta_{x_E}\star\delta_{x_E}=1$, with the odd powers then vanishing via $\tr_\star\delta_{x_E}=0$. We are left with:
\begin{align}
	\ln Z_{HS}(c) = \exp\left(\frac{N}{8}\sum_{n=1}^\infty \frac{(-1)^n c^{2n}}{n} \tr_\star\big(M_{ab}Y^aY^b\big)_\star^{2n}\right) \ . \label{eq:Z_spin_1_raw}
\end{align}
It remains to compute the star-product powers $\big(M_{ab}Y^aY^b\big)_\star^{2n}$. This can be done by recalling that quadratic functions such as $M_{ab}Y^aY^b$ are the generators of spacetime symmetries. In our case, $M_{ab}Y^aY^b$ generates an infinitesimal left-handed rotation around $x_E$. It is then easy to write its star-product exponential, which describes a finite rotation by angle $\theta$ \cite{Neiman:2017mel}:
\begin{align}
	\exp_\star\big(\theta M_{ab}Y^aY^b\big) = \frac{1}{\cos\theta}\exp\big(\tan\theta\,M_{ab}Y^aY^b\big) \ .
\end{align}
Setting $Y\to 0$, we read off the trace:
\begin{align}
	\tr_\star\exp_\star\big(\theta M_{ab}Y^aY^b\big) = \frac{1}{\cos\theta} \ .
\end{align}
Expanding both sides in powers of $\theta$, we conclude,:
\begin{align}
	\tr_\star\big(M_{ab}Y^aY^b\big)_\star^{2n} = (-1)^n E_{2n} \ ,
\end{align}
where $E_n$ are the Euler numbers. Plugging back into the partition function \eqref{eq:Z_spin_1_raw}, we get:
\begin{align}
	Z_{HS}(c) = \exp\left(\frac{N}{8}\sum_{n=1}^\infty \frac{c^{2n}}{n}\,E_{2n}\right) \ . \label{eq:Z_spin_1}
\end{align}
The power series in \eqref{eq:Z_spin_1} cannot be summed straightforwardly: it's non-convergent, due to $E_{n}$ growing faster than $n^{n}$. However, we can evaluate it using Borel summation. Let us denote the series and its Borel transform as:
\begin{equation}\label{eq:borel:transf}
	W(c) \equiv -\ln Z_{HS}(c) = -\frac{N}{8}\sum_{n=1}^\infty \frac{c^{2n}}{n}\,E_{2n} \ ; \quad \mathcal{B}W(c)= -\frac{N}{8}\sum_{n=1}^\infty \frac{c^{2n}}{n(2n)!}\,E_{2n} \ .
\end{equation}
It is straightforward to check that this new series is closely related to $1/\cosh c$, via:
\begin{equation}\label{eq:borel:rel1}
	c\frac{d}{dc}\mathcal{B}W(c) = \frac{N}{4}\left(1 - \frac{1}{\cosh c}\right)\ ,
\end{equation}
and thus:
\begin{equation}\label{eq:borel:transf_def2}
	\mathcal{B}W(c) = \frac{N}{4}\int_0^c \frac{db}{b}\left(1 - \frac{1}{\cosh b}\right) \ .
\end{equation}
We are now ready to perform the Borel summation, defined as:
\begin{equation}
	W(c) = \int_0^\infty e^{-t}\,\mathcal{B}W(tc)\,dt \ .
\end{equation}
Integrating by parts and using \eqref{eq:borel:rel1}, this becomes:
\begin{equation}
	W(c) = \frac{N}{4}\int_0^\infty \frac{e^{-t}}{t}\left(1 - \frac{1}{\cosh(tc)}\right)dt - \frac{N}{8}\left.\bigg(e^{-t}\,\mathcal{B}W(tc)\bigg)\right|_{t=0}^\infty \ . \label{eq:Borel_integral}
\end{equation}
The $t=0$ boundary term vanishes due to $\mathcal{B}W(0) = 0$, which is evident from \eqref{eq:borel:transf}. The $t=\infty$ boundary term also vanishes: it's easy to see from \eqref{eq:borel:transf_def2} that $\calB W(tc)$ grows with $t\to\infty$ only logarithmically, since the integrand in \eqref{eq:borel:transf_def2} is dominated by $1/b$. We are thus left with:
\begin{equation}\label{eq:borel:result}
	W(c) = \frac{N}{4}\int_0^\infty \frac{e^{-t}}{t}\left(1 - \frac{1}{\cosh(tc)}\right)dt = \frac{N}{2}\int_0^\infty \frac{e^{-t/|c|}\sinh^2\frac{t}{2}}{t\cosh t}\,dt \ , 
\end{equation}
where in the last expression we redefined the integration variable as $t\to t/|c|$. 

The integrals \eqref{eq:borel:result} are convergent for all real $c$. Moreover, it's manifest that $W(0)=0$ is a global minimum, i.e. $Z_{HS}(0) = 1$ is a global maximum, since the integrand in \eqref{eq:borel:result} increases monotonically with $|c|$. The behavior at $|c|\ll 1$ is easy to evaluate via the first expression in \eqref{eq:borel:result}. This gives:
\begin{align}
	W(c) \approx \frac{N}{8}c^2\int_0^\infty te^{-t}\,dt = \frac{N}{8}c^2 \quad \Rightarrow \quad Z_{HS}(c) \approx 1 - \frac{N}{8}c^2 \ , \label{eq:W_small_c}
\end{align}
as expected from the local-maximum behavior dictated by the 2-point functions of section \ref{sec:boundary:2_point}. At $|c|\gg 1$, it's easier to use the second expression in \eqref{eq:borel:result}. The $c$-dependence is then dominated by the large-$t$ tail of the integral, as:
\begin{align}
	W(c) \approx \const + \frac{N}{4}\int_{\const}^\infty \frac{e^{-t/|c|}}{t}\,dt = \const + \frac{N}{4}\ln|c| \quad \Rightarrow \quad Z_{HS}(c) \sim \frac{1}{|c|^{N/4}} \ . \label{eq:W_large_c}
\end{align}
Curiously, both the small-source asymptotic behavior \eqref{eq:W_small_c} and the large-source one \eqref{eq:W_large_c} are the same as for the spin-0 partition function \eqref{eq:Z_spin_0}. 

\subsection{Legendre (self-)duality of type-A and type-B} \label{sec:EAdS:Legendre}

Having established the HS-algebraic partition function \eqref{eq:Z_HS} and some of its basic properties, we return to one of the paper's main goals: to better understand the paradoxical equivalence of holography in the ``twice-smaller'' Euclidean AdS vs. the ``twice-larger'' Lorentzian de Sitter. As mentioned in the Introduction, we will do this by painting two pictures of the relationship between the type-A and type-B theories. In this section, we take up the Euclidean side of the story, where we make the claim that type-A and type-B can be interpreted as \emph{the same theory with opposite boundary data}. In other words, we claim that the \emph{type-A and type-B partition functions are related by a Legendre transform}. This was speculated by us previously in \cite{Lang:2024dkt}, starting from the somewhat different context of $\calN=2$ supersymmetric Euclidean AdS/CFT. Here, we make it more concrete, in the specific context of dS/CFT with no supersymmetry.

First, we must stress that our ``type-A = type-B'' claim is at odds with the usual understanding, where the bulk interactions of type-A and type-B are simply different. In fact, the claim is made possible only by ``taking seriously'' the HS-algebraic partition function \eqref{eq:Z_HS}, which, as we recall, differs from the standard CFT one \eqref{eq:Z_local} beyond the 2-point level. With that disclaimer, let us now present our evidence. 

\subsubsection{The span of the Euclidean boundary-bulk propagators} \label{sec:EAdS:Legendre:propagators}

We start with the standard observation \cite{Didenko:2013bj,Lang:2024dkt} that the type-A boundary-bulk propagators \eqref{eq:type_A_propagator} and their type-B counterparts \eqref{eq:type_B_propagator} have opposite boundary conditions. Specifically, the spin-0 type-A propagator satisfies boundary conditions with conformal weight $\Delta=2$, while the type-B propagator satisfies boundary conditions with weight $\Delta=1$. For nonzero spins $s>0$, the type-A propagators satisfy magnetic boundary data, while the type-B propagators satisfy electric ones. We note that in the standard picture of the Vasiliev equations, the difference in boundary conditions for $s>0$ arises from different conventions for the relation between the curvature master-field $C(x;Y)$ and the connection master-field $\Omega(x;Y)$; in terms of $\Omega(x;Y)$, the boundary conditions are always magnetic. However, in the present context, we stubbornly insist on treating $C(x;Y)$ as the \emph{same} master field for both type-A and type-B, which simply happens to take different values for the type-A propagators \eqref{eq:type_A_propagator} vs. the type-B ones \eqref{eq:type_B_propagator}. The boundary conditions are then indeed opposite for all spins.

Our next observation is that, in dS/CFT, the type-A and type-B propagators \emph{span the same space of field configurations in Euclidean AdS}. This statement is new, and it's specific to the spacetime signature and the sign of the cosmological constant. In Lorentzian AdS, propagators with opposite boundary conditions can be taken to span separate spaces of bulk field configurations. In Euclidean AdS, bulk regularity forces linear relations between the opposite types of propagators, so that they span the same space of configurations, at least \emph{in the complex sense}. However, as we saw in \cite{Lang:2024dkt}, in Euclidean AdS/CFT the type-A and type-B propagators do not span the same \emph{real} space: there is always a factor of $i$ between one set and the other. But, as we saw in section \ref{sec:boundary:type_B_dS}, to incorporate type-B correctly into dS/CFT, one must insert an overall factor of $i$ into the boundary action \eqref{eq:dS_CFT_type_B} vs. \eqref{eq:EAdS_CFT_type_B}, which translates into a factor of $1/i$ on all the type-B boundary-bulk propagators. As a result, in dS/CFT, the type-A propagators \eqref{eq:type_A_propagator} and type-B propagators \eqref{eq:type_B_propagator} \emph{span the same real space}. Indeed, they satisfy the same reality condition \eqref{eq:propagators_reality} in Euclidean AdS. As a result, when we presented the basis \eqref{eq:C_Psi}-\eqref{eq:F_Psi},\eqref{eq:Psi_reality} for the boundary-bulk propagators integrated against boundary sources, we never needed to specify whether we're talking about type-A or type-B.

In summary, the type-A and type-B boundary-bulk propagators span the same space of configurations in Euclidean AdS, but have opposite boundary conditions. This motivates our claim that the type-A and type-B theories may be viewed as the same, up to a Legendre transform that flips the boundary data between one type ($\Delta=2$ for $s=0$, magnetic for $s>0$) and the other ($\Delta=1$ for $s=0$, electric for $s>0$). Let us now test this claim at the level of the partition function.

\subsubsection{Legendre (self-)duality of the HS-algebraic partition function}

We start with yet another standard observation \cite{Didenko:2013bj,Lang:2024dkt}: for both type-A and type-B, the HS-algebraic correlator formula \eqref{eq:correlator_dS_type_A}, and thus also the formula \eqref{eq:Z_HS} for the HS-algebraic partition $Z_{HS}$, are the same. As a result, when we computed the partition function $Z_{HS}$ for various source harmonics in sections \ref{sec:EAdS:partition_function}-\ref{sec:EAdS:spin_1}, we again never needed to specify whether we're talking about type-A or type-B. Once again, this is all due to the difference between $Z_{HS}$ and the standard CFT partition function \eqref{eq:Z_local}. 

Now, the fact that type-A and type-B are governed by the same partition function \eqref{eq:Z_HS} already suggests that at this (twistorial, HS-algebraic) level, they are the same theory. However, as discussed above in section  \ref{sec:EAdS:Legendre:propagators}, this shouldn't quite be our expectation: we should expect the type-A and type-B partition functions to be not equal, but rather related by a Legendre transform. To resolve this tension, and present our final piece of evidence for the ``type-A = type-B'' claim, we make the following observation: \emph{the partition function of HS Gravity is equal to its own Legendre transform}. This statement is very basic, but appears to be new. Let us now define and demonstrate it.

We start from the HS-algebraic partition function \eqref{eq:Z_HS}. We define an ``effective action'' $W[F(Y)]$ by taking its logarithm:
\begin{align}
	W[F(Y)] \equiv -\ln Z_{HS}[F(Y)] \ , \label{eq:W}
\end{align}
We then define the canonical conjugate $G(Y)$ to the source $F(Y)$, by varying this effective action:
\begin{align}
	\delta W \equiv \tr_\star\big(G(Y)\star\delta F(Y)\big) \ . \label{eq:G}
\end{align}
Note that the definitions \eqref{eq:W}-\eqref{eq:G} are Euclidean-style, in that they don't contain factors of $i$. One can switch to Lorentzian-style definitions by replacing $W$ with $-iW$ everywhere; this won't affect the result below. Finally, we define the Legendre-transformed effective action as:
\begin{align}
	\tilde W \equiv \tr_\star\big(G(Y)\star F(Y)\big) - W \ . \label{eq:tilde_W}
\end{align}
Our statement then is that, for the specific partition function \eqref{eq:Z_HS} of HS gravity, the function $\tilde W[G(Y)]$ is the same as the original $W[F(Y)]$. Let us now show that this is indeed the case. From the partition function \eqref{eq:Z_HS}, we get:
\begin{align}
	W[F(Y)] = \tr_\star\left(\frac{N}{4}\ln_\star\left[1-\sqrt{\frac{4}{N}}\,F(Y)\right] + \sqrt{\frac{N}{4}}\,F(Y) \right) \ . \label{eq:W_specific}
\end{align}
To obtain the canonical conjugate $G(Y)$ as defined by \eqref{eq:G}, we simply differentiate the star-product function appearing in \eqref{eq:W_specific}:
\begin{align}
	G(Y) = -\left(\frac{F(Y)}{1 - \sqrt{4/N}\,F(Y)} \right)_\star \ , \label{eq:G_specific}
\end{align}
where the $(\dots)_\star$ subscript again indicates that all the products in the expression's Taylor expansion should be understood as star-products. The relation \eqref{eq:G_specific} is its own inverse, i.e. we can extract:
\begin{align}
	F(Y) = -\left(\frac{G(Y)}{1 - \sqrt{4/N}\,G(Y)} \right)_\star \ .
\end{align}
Plugging everything into \eqref{eq:tilde_W}, we obtain:
\begin{align}
	\begin{split}
   	   \tilde W &= -\tr_\star\left(\frac{N}{4}\ln_\star\left[1-\sqrt{\frac{4}{N}}\,F(Y)\right] + \sqrt{\frac{N}{4}}\left(\frac{F(Y)}{1 - \sqrt{4/N}\,F(Y)} \right)_\star \right) \\
   	     &= \tr_\star\left(\frac{N}{4}\ln_\star\left[1-\sqrt{\frac{4}{N}}\,G(Y)\right] + \sqrt{\frac{N}{4}}\,G(Y) \right) \ ,
	\end{split}
\end{align}
i.e. the Legendre transform $\tilde W[G(Y)]$ is indeed the same as the original $W[F(Y)]$. 

Note that in the above calculation, the star-products and traces were just ``along for the ride''. Indeed, our result is a simple consequence of the fact that the function $W(x) = a^2\ln(1-x/a) + ax$ is equal to its own Legendre transform. This fact is well-known to mathematicians, and is rather peculiar: no other such elementary function is known to exist (aside from the ``free-field limit'' $W(x) = -x^2/2$ at $a\to\infty$). Thus, it seems truly non-trivial that HS Gravity possesses this property. We are compelled to take this observation seriously, and with it the strange picture of ``type-A = type-B''.

\section{Type-A and type-B as parallel ``elliptic de Sitter'' worlds} \label{sec:elliptic_dS}

In section \ref{sec:EAdS:Legendre} above, we painted a picture where type-A and type-B are the same theory in Euclidean AdS, merely with opposite choices of boundary conditions. In this section, we set out to paint a complementary picture, in which type-A and type-B describe two ``parallel worlds'' in the real de Sitter spacetime. 

We start by recalling the two opposite sets of boundary conditions associated with free CFT's on the boundary: the type-A conditions ($\Delta=2$ for $s=0$, magnetic for $s>0$), and the type-B conditions ($\Delta=1$ for $s=0$, electric for $s>0$). In Euclidean AdS, both of these sets of boundary data are always present, and are related via the Legendre transform. The only thing distinguishing one or the other is which one enters as the argument of the partition function. However, in Lorentzian de Sitter, the two sets of boundary data can become truly independent: there are real solutions to the free bulk field equations that have only one set of boundary data non-vanishing, or only the other. These solutions have been studied extensively in \cite{Neiman:2014npa,Halpern:2015zia}. Their key property is \emph{antipodal symmetry}: under the $dS_4$ antipodal map $x^\mu\to-x^\mu$, which interchanges $\scri^+\leftrightarrow\scri^-$ and $EAdS_4^+\leftrightarrow EAdS^-_4$, the solutions with purely type-A boundary data are \emph{even}, while those with purely type-B boundary data are \emph{odd}. 

There is a subtlety here. Naively, the boundary-bulk propagators \eqref{eq:type_A_propagator}-\eqref{eq:type_B_propagator} have the opposite antipodal symmetries from those stated above: odd for type-A, e.g. the scalar propagator $\sim 1/(\ell\cdot x)$, and even for type-B, e.g. the scalar propagator $\sim 1/(\ell\cdot x)^2$. The issue is again the choice of $i\varepsilon$ prescription in the denominators. The propagators with purely type-A or purely type-B boundary data respectively arise from taking the \emph{difference} of the $\pm i\varepsilon$ prescriptions. This makes delta-function-like propagators, such as $\sim\delta(\ell\cdot x)$ for the type-A scalar, and $\sim\delta'(\ell\cdot x)$ for the type-B scalar. These satisfy our stated antipodal symmetry: even for type-A, odd for type-B.

Imposing distinct antipodal symmetries on the type-A and type-B fields essentially places them in two ``parallel worlds'' within de Sitter space. The geometric version of this statement is to consider the so-called ``elliptic'' de Sitter space $dS_4/\bbZ_2$ \cite{Parikh:2002py,Neiman:2014npa,Halpern:2015zia}, i.e. the orbifold of $dS_4$ under the antipodal map $x^\mu\leftrightarrow-x^\mu$. Antipodally even and odd fields then belong to two different bundles on $dS_4/\bbZ_2$: the trivial bundle for the even case, and a Mobius-band-like bundle for the odd case. In this sense, the type-A and type-B sectors are not just two separate sets of fields as in the standard picture of Vasiliev \cite{Vasiliev:1990en,Vasiliev:1992av}, but they in fact live on two separate geometries. Furthermore, from the point of view of the bulk HS Gravity, the separation into ``type-A = antipodally-even'' and ``type-B = antipodally-odd'' worlds should survive the upgrade from free fields to interacting ones. Indeed, as discussed in \cite{Neiman:2014npa}, the antipodal map is a discrete symmetry of the CT type, which is equivalent under CPT to parity P. As a consequence, parity-preserving interactions also preserve antipodal symmetry, provided that parity-even fields are antipodally even, and parity-odd fields are antipodally odd. Since the interactions of type-A/type-B HS Gravity are parity-preserving, with the curvature master field $C(x;Y)$ parity-even/parity-odd respectively, we conclude that the interactions should indeed preserve our prescribed antipodal symmetries.

Let us now return to the holographic partition function \eqref{eq:Z_HS}, whose arguments are the integrated boundary-bulk propagators \eqref{eq:integrated_C}-\eqref{eq:integrated_F}. We can replace these integrated propagators (in the type-A or type-B theory respectively) by their versions with purely type-A or type-B boundary data. Since the boundary data is in one-to-one correspondence with antipodal symmetry, this simply means projecting onto the \emph{antipodally even or odd part}, respectively, of the fields \eqref{eq:integrated_C}-\eqref{eq:integrated_F}. Recalling eqs. \eqref{eq:C_antipodal}-\eqref{eq:F_antipodal}, this can be implemented in HS algebra using the antipodally even/odd projectors:
\begin{align}
	P_\pm(Y) \equiv \frac{1}{2}\big(1 \pm \delta(Y)\big) \ ; \quad P_\pm(Y)\star P_\pm(Y) = P_\pm(Y) \ ; \quad P_\pm(Y)\star P_\mp(Y) = 0 \ .
\end{align}
Thus, for e.g. the type-A theory, we take the antipodally-even part of the integrated propagators \eqref{eq:C_antipodal}-\eqref{eq:F_antipodal}:
\begin{align}
	F_A(Y) \equiv F(Y)\star P_+(Y) \ , \label{eq:F_A}
\end{align}
which satisfies by construction the antipodal symmetry:
\begin{align}
	F_A(Y)\star\delta(Y) = F_A(Y) \ . \label{eq:F_A_even}
\end{align}
Let us now try to plug this, instead of the positive-frequency $F(Y)$, into the HS-algebraic partition function formula \eqref{eq:Z_HS}. Since the projectors $P_\pm(Y)$ commute with any even (i.e. integer-spin) function, this becomes simply:
\begin{align}
	Z_{HS}[F_A(Y)] = \exp\left(\frac{N}{4} \sum_{n=2}^\infty \frac{1}{n}\left(\frac{4}{N}\right)^{n/2}\tr_\star\Big(\underbrace{F(Y)\star\ldots\star F(Y)}_{n\text{ factors}}{}\star P_+(Y)\Big) \right) \ . \label{eq:Z_A}
\end{align}
For $F(Y)$ consisting of finitely many harmonics from the basis \eqref{eq:C_Psi}-\eqref{eq:F_Psi}, the computation of \eqref{eq:Z_A} follows the same general pattern as described at the end of section \ref{sec:EAdS:partition_function} for the original  \eqref{eq:Z_HS}. At odd orders $n$, the powers $\big(F(Y)\big)^n_\star$ contain a factor of $\delta_{x_E}(Y)$ acted on by polynomials in $Y$ and $\del/\del Y$. Multiplication by the $\delta(Y)$ inside the $P_+(Y)$ projector replaces the $\delta_{x_E}$ by $\delta_{-x_E}$. We take the trace $\tr_\star$ of all such terms to vanish, by the same symmetry arguments cited in section \ref{sec:EAdS:partition_function}. At even orders $n$, the powers $\big(F(Y)\big)^n_\star$ are polynomials in $Y$. Multiplication by $\delta(Y)$ turns these into polynomials in $\del/\del Y$ acting on $\delta(Y)$. By the $SO(4)$ symmetry of rotations around $x_E$, the trace of all nonzero powers in these polynomials must vanish. As a result, the exponent in \eqref{eq:Z_A} takes the same value as the original \eqref{eq:Z_HS}, but with the final factors of $\tr_\star 1 = 1$ replaced by $\tr_\star P_+(Y) = \frac{1}{2}(1 + \tr_\star\delta(Y))$:
\begin{align}
	Z_{HS}[F_A(Y)] = \big(Z_{HS}[F(Y)]\big)^{(1 + \tr_\star\delta(Y))/2} \ . \label{eq:Z_A_final}
\end{align}
For the type-B theory, an analogous discussion holds, and \eqref{eq:F_A}-\eqref{eq:Z_A_final} are replaced by:
\begin{align}
	F_B(Y) &\equiv F(Y)\star P_-(Y) \ ; \\
	F_B(Y)\star\delta(Y) &= -F_B(Y) \ ; \label{eq:F_B_odd} \\
	\begin{split}
  	  Z_{HS}[F_B(Y)] &= \exp\left(\frac{N}{4} \sum_{n=2}^\infty \frac{1}{n}\left(\frac{4}{N}\right)^{n/2}\tr_\star\Big(\underbrace{F(Y)\star\ldots\star F(Y)}_{n\text{ factors}}{}\star P_-(Y)\Big) \right) \\
  	    &= \big(Z_{HS}[F(Y)]\big)^{(1 - \tr_\star\delta(Y))/2} \ .
	\end{split} \label{eq:Z_B}
\end{align}
Though we are tempted to set $\tr_\star\delta(Y)$ to zero, this doesn't seem to be required by any symmetry. Finally, we can consider sources of the type-A and type-B types simultaneously (each corresponding to a separate positive-frequency source $F(Y)$). Such sources can be described by a general integer-spin function $F_{\text{tot}}(Y)$, with $F_A(Y)$ and $F_B(Y)$ as its antipodally-even and antipodally-odd parts:
\begin{align}
	F_{\text{tot}}(Y) = F_A(Y) + F_B(Y) \ .
\end{align}
Plugging \emph{this} into the HS-algebraic partition function formula \eqref{eq:Z_HS}, we find trivial factorization, as expected for ``parallel worlds'':
\begin{align}
	Z_{HS}[F_{\text{tot}}(Y)] = Z_{HS}[F_A(Y)]\,Z_{HS}[F_B(Y)] \ . \label{eq:Z_tot}
\end{align}
The reason for this factorization is that any mixed product of the form $F_A\star F_B$ or $F_B\star F_A$ vanishes, due to the orthogonality of projectors $P_+(Y)\star P_-(Y) = P_-(Y)\star P_+(Y) = 0$.

The antipodally-symmetric sources $F_{A/B}(Y)$, and thus also $F_{\text{tot}}(Y)$ inherit the reality property of the positive-frequency sources \eqref{eq:integrated_reality}:
\begin{align}
	F_{A/B}^*(Y) = F_{A/B}(Y) \ ; \quad F_{\text{tot}}^*(Y) = F_{\text{tot}}(Y) \ . \label{eq:combined_reality}
\end{align} 
When combined with the antipodal symmetry \eqref{eq:F_A_even},\eqref{eq:F_B_odd}, this reproduces our $dS_4$ reality conditions \eqref{eq:reality_star}, with $\eta = \pm 1$ respectively for the type-A and type-B sector.

We still lack a clear interpretation for the partition functions \eqref{eq:Z_A_final},\eqref{eq:Z_B} that use antipodally-symmetric rather than positive-frequency fields as their arguments. Nevertheless, we suspect that this construction will have some importance. One argument in favor is that, unlike the positive-frequency condition (or the condition of regularity on $EAdS_4^+$), the antipodal symmetry conditions $F_{A/B}(Y)\star\delta(Y) = \pm F_{A/B}(Y)$ have a simple expression within HS algebra. 

\section{Discussion} \label{sec:discuss}

In this paper, we expanded Higher-Spin dS/CFT to odd spins and to type-B. In this process, we verified the correct real forms \cite{Chang:2013afa} of the Vasiliev equations for dS/CFT, and constructed the correct real forms of the boundary vector models, as well as an HS-algebraic construction of the Hilbert space for a de Sitter HS multiplet. We then used HS algebra to flesh out both sides of the tension between Euclidean AdS and the twice-larger de Sitter: one picture where type-A and type-B are the same theory in Euclidean AdS, and another where they live in two ``parallel'' antipodally-symmetric universes within de Sitter.

Several open questions present themselves. It would be interesting to further study the supersymmetry of the single-particle Hilbert space from section \ref{sec:Hermitian:symmetry}. We saw that the Hermitian conjugates of the fermionic generators take us outside of the original HS superalgebra. If we nevertheless include them, what is the resulting superalgebra? How close does it come to covering all possible transformations on the Hilbert space? Since the fermionic generators are chiral, is the superalgebra of section \ref{sec:Hermitian:symmetry} perhaps relevant for Chiral HS Gravity \cite{Ponomarev:2016lrm,Skvortsov:2018jea,Skvortsov:2018uru,Didenko:2022qga}? For existing work on supersymmetric Chiral HS Gravity, see \cite{Tsulaia:2022csz}.

On the topic of the Hilbert space in de Sitter, there is an understanding within GR (already at the free-field level \cite{Moncrief:1979bg,Higuchi:1991tk,Higuchi:1991tm}) that one must average all states under the global $SO(1,4)$ symmetry. It is unclear whether the same is true for HS gravity and HS symmetry (in particular, see \cite{Neiman:2015wma,Neiman:2022enh} for formulations of HS gravity that live on a fixed geometry, making the role of diffeomorphisms very different from that in GR). If so, then the single-particle Hilbert space of section \ref{sec:Hermitian} will trivialize: averaging over $SO(1,4)$ will leave only one state for each helicity, and averaging over the whole HS group will leave only one state overall. In that case, a proper discussion of the theory's Hilbert space can only begin at the multi-particle level, as was done for GR in \cite{Higuchi:1991tm}.

Turning to the boundary CFT's of section \ref{sec:boundary}, we saw that the dS/CFT boundary fermion action \eqref{eq:dS_CFT_type_B} is real, unlike that of Euclidean AdS/CFT \eqref{eq:EAdS_CFT_type_B}. Is that a physical principle in dS/CFT, akin to the unitarity requirement that the action must be real in Lorentzian theories?

It would be nice to extend the partition function calculation of section \ref{sec:EAdS:spin_1} to higher harmonics and higher spins. In particular, how universal is the strong-source asymptotics we observed for the lowest spin-0 and spin-1 harmonics? 

In section \ref{sec:EAdS:Legendre}, we observed that the HS partition function is self-dual under the Legendre transform, which proved important for our discussion of the tension between Euclidean AdS and de Sitter. Can this Legendre-self-duality be elevated to a principle of dS/CFT? Can it, together with HS symmetry and perhaps some assumptions on e.g. strong-source asymptotics, uniquely fix the partition function?

The most conceptually raw part of our paper is probably section \ref{sec:elliptic_dS}. Is it in fact sensible to evaluate the partition function on de Sitter solutions with pure boundary data of a single type, instead of positive-frequency solutions that are regular on $EAdS_4^+$? What is the physical meaning of such partition functions \eqref{eq:Z_A_final},\eqref{eq:Z_B},\eqref{eq:Z_tot}? Note that this question presents itself only due to the unfolded/twistorial formalism of HS theory, where it's more natural to work with entire linearized bulk solutions than with boundary data as such. Finally, we saw (just like in \cite{Lang:2024dkt} in a different context) that the antipodal-map operator $F(Y)\to F(Y)\star\delta(Y)$ in section \ref{sec:elliptic_dS} can be equated with the sign parameter/operator $\eta$ that distinguishes type-A from type-B in the Vasiliev equations \eqref{eq:dW}-\eqref{eq:S_S}. This highlights the need to better understand the antipodal map within the embedding-space formulation \cite{Neiman:2015wma} of the Vasiliev equations. In particular, we made a general argument in section \ref{sec:elliptic_dS} that type-A and type-B can be consistently formulated on $dS_4/\bbZ_2$ also at the non-linear level. Can this statement be made precise at the level of the Vasiliev equations, like we did for the reality conditions in section \ref{sec:reality:Vasiliev}?

Finally, we note that the recent paper \cite{Anninos:2026hia} has proposed a separate sense in which de Sitter is a ``twice-larger'' version of Euclidean AdS. There, the authors consider \emph{Euclidean} de Sitter, i.e. the $S_4$ that arises from analytically continuing the de Sitter static patch. A conformal holographic theory is then defined on an equatorial $S_3$, which becomes analogous to the Euclidean AdS boundary, and is glued to a second copy of itself in order to complete the $S_4$ path integral. While this construction is different in many details from our discussion of the ``doubled nature'' of de Sitter, it would be interesting to look for common ground between the two.

\section*{Acknowledgements}

We are grateful to Nick Dorey and Mirian Tsulaia for discussions, and to Chi-Ming Chang for an email exchange. This work was supported by the Quantum Gravity Unit of the Okinawa Institute of Science and Technology Graduate University (OIST).

\end{document}